\documentclass[pre,aps,10pt,superscriptaddress,amsmath,amssymb,showpacs,twocolumn,floatfix]{revtex4-1}

\usepackage[unicode]{hyperref}

\usepackage{bm}
\usepackage{graphicx}
\usepackage{epstopdf}

\def\be{\begin{equation}}
\def\ee{\end{equation}}
\def\vp{\varphi}

 \DeclareMathOperator{\Tr}{Tr}
 \DeclareMathOperator{\tr}{tr}

 \DeclareMathOperator{\arccosh}{arccosh}

\def\br{\mathbf{r}}

\def\bq{\mathbf{q}}

\newcommand{\corr}[1]{\langle #1\rangle}
\newcommand{\ccorr}[1]{\langle\langle #1\rangle\rangle}

\def\Re{\mathop\mathrm{Re}}
\def\Im{\mathop\mathrm{Im}}

\begin{document}

\title{Subgap states in disordered superconductors with strong magnetic impurities}

\author{Yakov~V.\ Fominov}
\affiliation{L.~D.\ Landau Institute for Theoretical Physics RAS, 142432 Chernogolovka, Russia}
\affiliation{Moscow Institute of Physics and Technology, 141700 Dolgoprudny, Russia}

\author{Mikhail~A.\ Skvortsov}
\affiliation{Skolkovo Institute of Science and Technology, 143026 Skolkovo, Russia}
\affiliation{L.~D.\ Landau Institute for Theoretical Physics RAS, 142432 Chernogolovka, Russia}
\affiliation{Moscow Institute of Physics and Technology, 141700 Dolgoprudny, Russia}

\date{25 April 2016}

\begin{abstract}
We study the density of states (DOS) in diffusive superconductors
with pointlike magnetic impurities of arbitrary strength described by the Poissonian statistics.
The mean-field theory predicts a nontrivial structure of the DOS with the continuum of quasiparticle states and (possibly) the impurity band.
In this approximation,
all the spectral edges are hard,
marking distinct boundaries between spectral regions of finite and zero DOS. Considering instantons in the replica sigma-model technique, we calculate the
average
DOS beyond the mean-field level
and determine the smearing of the spectral edges due
interplay of fluctuations of potential and nonpotential disorder.
The latter, represented by inhomogeneity in the concentration of magnetic impurities, affects the subgap DOS in two ways:
via fluctuations of the pair-breaking strength
and via induced fluctuations of the order parameter. In limiting cases, we reproduce previously reported results for the subgap DOS in disordered superconductors with strong magnetic impurities.
\end{abstract}

\pacs{74.25.Jb, 74.81.-g, 75.20.Hr, 73.22.-f}


%
%
%

\maketitle


\section{Introduction}

The influence of local inhomogeneities on the density of states (DOS) in superconductors depends on the nature of disorder. In $s$-wave superconductors,
potential impurities do not change the BCS DOS \cite{AGdirty,Anderson}, while magnetic impurities cut the coherence peak and suppress the superconducting gap. In the simplest model of weak magnetic impurities (Born limit) studied by Abrikosov and Gor'kov (AG) \cite{AG}, the spectral gap $E_g$ is reduced compared to the order parameter $\Delta$, and the BCS edge singularity $\rho(E)\propto(E-\Delta)^{-1/2}$ is replaced by the square-root vanishing behavior $\rho(E)\propto(E-E_g)^{1/2}$.

The effect of magnetic impurities on the superconducting state becomes more fascinating beyond the Born limit.
In this case, a single magnetic impurity produces a localized state inside the BCS gap \cite{Yu,Soda,Shiba,Rusinov}, which
can be visualized experimentally \cite{Yazdani1997,Ji2008,Franke2011,Roditchev2015}.
At finite concentration of magnetic impurities, the states localized on different impurities overlap and form an impurity band. As the concentration grows, the band becomes wider. If it merges with the continuum of quasiparticle states, then the
AG-like regime is realized. Alternatively, the impurity band can touch
the Fermi energy ($E=0$)
before merging with the continuum (see Ref.~\cite{Balatsky-review} for a review).
Although various structures of the DOS with several spectral edges can be realized depending on the parameters of magnetic disorder (strength of individual impurities and their concentration), a general feature of the mean-field
results \cite{AG,Yu,Soda,Shiba,Rusinov}
is that all the gaps $E_{gi}$ in the spectrum remain hard, sharply dividing energy regions with
zero and finite [with $\rho(E)\propto|E-E_{gi}|^{1/2}$] DOS.

The square-root vanishing of the DOS is not specific to superconductors with magnetic disorder.
The same qualitative behavior is observed, e.g., in the mean-field treatment of proximity-coupled normal-superconducting (NS) structures \cite{McMillan,GolubovKupriyanov}, in the model of a random Cooper-channel interaction constant \cite{LO_1971}, in the random-matrix theory (Wigner semicircle) \cite{Wigner,Mehta}, for imbalanced vacancies in graphene \cite{vacancies}, etc.

Existence of the sharp spectral edge is an artefact of the mean-field approximation.
The exact treatment reveals a tail of the subgap states formed in the classically gapped region.
The physical origin of these states is related to fluctuations,
when some rear disorder configurations, missed on the mean-field level,
lead to local shifts of the spectral edges. Averaging over fluctuations then results in the spatially homogeneous nonzero subgap DOS.

Fluctuation smearing of the gap edge was first considered by Larkin and Ovchinnikov (LO) in the model
of a diffusive superconductor with
short-range disorder in the Cooper-channel constant \cite{LO_1971}. They described formation of subgap states in the language of optimal fluctuations of the order-parameter field $\Delta(\br)$, generalizing the method originally developed in the studies of doped semiconductors (for a review, see Ref.~\cite{LGP}).
The resulting average DOS decays with the stretched-exponential law as a function of $|E-E_g|$, with the width of the tail determined by the magnitude of $\Delta(\br)$ fluctuations.

A different approach to the description of the subgap states
in diffusive superconductors
was elaborated in early 2000s by Simons and co-authors \cite{LamacraftSimons,MeyerSimons2001,MS}, who considered instanton configurations in the nonlinear sigma-model formalism.
They also obtained a stretched-exponential decay of the average DOS as a function of the distance $|E-E_g|$.
In contrast to the LO theory, the smallness of this effect is controlled by the large normal-state conductance, $g$, rather than by the magnitude of $\Delta(\br)$ fluctuations,
indicating that the tail obtained is determined solely by fluctuations of potential disorder.
The same type of instanton was shown to describe the smearing of the minigap in SNS junctions \cite{OSF01}.
Physically, the subgap states obtained in Refs.~\cite{LamacraftSimons,MeyerSimons2001,MS,OSF01} are due to mesoscopic fluctuations
originating from the randomness of potential disorder.
In this respect, they resemble the states beyond the Wigner semicircle in the random-matrix theory \cite{TracyWidom}. This analogy was exploited in Refs.~\cite{Narozhny,Vavilov}, where tail formation in zero-dimensional superconducting systems was studied. The results by Simons \emph{et al.} \cite{LamacraftSimons,MeyerSimons2001,MS} can then be considered as a direct generalization of previous random-matrix results to nonzero dimensionalities.

The apparent discrepancy between the results of Larkin and Ovchinnikov \cite{LO_1971} and Meyer and Simons \cite{MeyerSimons2001} for the random-coupling model
was recently resolved in Ref.~\cite{SF}, where it was demonstrated that the two different regimes correspond to different limits of the same instanton solution.
Sufficiently close to the gap edge, at small $|E-E_g|$, fluctuations of $\Delta(\br)$ (nonpotential disorder) are more important, and the subgap DOS is described by the LO theory.
In the far asymptotics realized at sufficiently large $|E-E_g|$, the DOS tail is determined by more efficient mesoscopic fluctuations (potential disorder).
Mathematically, interplay of these two different physical sources of disorder manifests itself as the competition between two types of nonlinearities in the instanton equations.

An important feature of disorder-induced gap smearing is its large degree of universality in the vicinity of $E_g$, where any type of nonpotential disorder can be mapped onto
an effective random order parameter (ROP) model \cite{SF}.
This mapping should be understood in a sense that the DOS smearing in the original problem is equivalent to the DOS smearing in an artificial model, where $\Delta(\br)$ is the only fluctuating quantity (even though the order parameter may not fluctuate in the original problem).
The parameters of the initial quenched inhomogeneity are then encoded in the correlation function $f(\bq)=\ccorr{\Delta\Delta}_\bq$ in the artificial ROP model to be determined for a particular problem.
The ROP model thus provides a universal account
of the subgap DOS, describing the interplay of fluctuations due to potential and nonpotential disorder.

The problem
with infinitesimally weak magnetic impurities
was reduced to the ROP model in Ref.~\cite{SF}, where the leading source of nonpotential fluctuations was identified as disorder in local magnetization (triplet sector).
Due to the nonlinearity of the Usadel equation, this disorder translates into fluctuations in the singlet sector, that are equivalent to an emergent inhomogeneity of the effective order parameter.
This mechanism (referred to as direct in Ref.~\cite{SF}) leads to a sufficiently small ROP correlation function, so it is possible to have a situation when the LO regime is unobservable and the full tail is due to mesoscopic fluctuations as described by Lamacraft and Simons \cite{LamacraftSimons}.

Subgap states due to magnetic impurities in otherwise clean superconductors were considered in Refs.~\cite{BalatskyTrugman,Shytov}. We are interested in the opposite situation, in which the underlying electron dynamics (in the absence of magnetic impurities) is diffusive due to potential scattering.

\begin{figure}[t]
\centering{\includegraphics[scale=1]{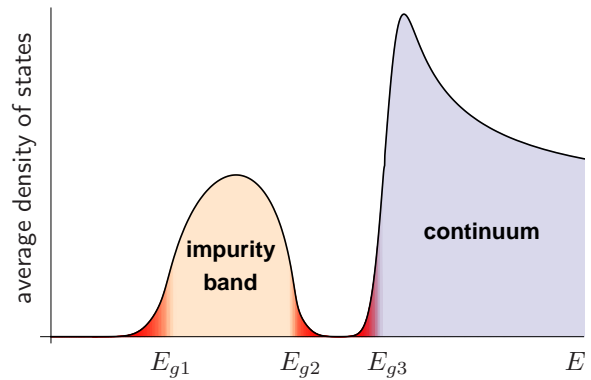}}
\caption{
Schematic form of the average DOS. Mean-field hard gaps at $E_{gi}$
are smeared due to fluctuations of the concentration of magnetic impurities
and/or mesoscopic fluctuations of potential disorder.}
\label{F:dos-smeared}
\end{figure}

The purpose of the present paper is to extend the approach of Ref.~\cite{SF} to the case of strong magnetic impurities
and to quantitatively describe fluctuation smearing of the gap edges (see Fig.~\ref{F:dos-smeared}).
In the vicinity of the mean-field edge, we reduce the problem to the ROP model and calculate the effective correlation function $f(0)$.
The principal difference from the Born limit is that now the primary source of disorder is due to fluctuations of the concentration of magnetic impurities
which leads to a larger correlation function $f(0)$
in the ROP model, as it does not require excitations of the triplet modes.
As a consequence, the previous results by Marchetti and Simons \cite{MS} describe only the far asymptotics of the DOS tails due to mesoscopic fluctuations, whereas the main asymptotics is given by the LO-type expression arising due to Poissonian fluctuations of magnetic disorder.
The importance of the Poissonian statistics of magnetic impurities was realized by Silva and Ioffe \cite{Silva}, who found the main asymptotics of the subgap DOS in the case of weak impurities (close to the Born limit). We reproduce their result in the corresponding limiting case.

The paper is organized as follows. In Sec.~\ref{S:modelresults}, we formulate the model and discuss the main results. In Sec.~\ref{S:fieldtheory}, we formulate our field-theoretical approach, underlining the procedure of averaging over Poissonian statistics of magnetic impurities. Section~\ref{S:RSB} is devoted to description of the replica-symmetry-breaking instanton solution, responsible for the subgap DOS. We also map our problem to the ROP model. In Sec.~\ref{S:Limiting}, the developed approach is applied to several limiting cases. Possibility of experimental observation of the predicted DOS tails is discussed in Sec.~\ref{S:discussion}. Finally, we present our conclusions in Sec.~\ref{S:concl}. Some technical details are presented in Appendices.

Throughout the paper we employ the units with $\hbar = k_B = 1$.

\section{Model and results}
\label{S:modelresults}

\subsection{Model of magnetic impurities}
\label{SS:Model}

We consider a dirty $s$-wave superconductor with both potential and magnetic disorder.
Scattering on potential impurities preserving the electron spin is assumed
to be the dominant mechanism of momentum relaxation.
On time scales larger than the elastic mean free time $\tau$, electron motion
becomes diffusive with the diffusion constant $D$.
A much weaker magnetic (spin-flip) scattering is described by the Hamiltonian
\begin{equation}
  H_\text{mag}
  =
  J
  \int d^3\mathbf r \, \hat\psi^\dagger_\sigma(\mathbf r) \mathbf S(\mathbf r) \boldsymbol\sigma \hat\psi_\sigma(\mathbf r) .
\end{equation}

We make the same assumptions about the magnetic disorder as in Refs.~\cite{Yu,Soda,Shiba,Rusinov,MS}:
(a)~it is classical with the spin density
\begin{equation}
\label{spin-density}
\mathbf S(\mathbf r) = \sum_i \delta(\mathbf r-\mathbf r_i) \mathbf S_i,
\end{equation}
where the points $\mathbf r_i$ have a Poisson distribution, and
(b)~spins of different magnetic impurities are statistically independent and the distribution over orientations is uniform, $P(\{ \mathbf S_i\}) = \prod_i \delta( \mathbf S_i^2- S^2)$.

Magnetic impurities are characterized by the two dimensionless parameters:
$0<\mu\leqslant1$ and $0<\eta$.
The parameter of ``unitarity'' $\mu$, defined as
\begin{equation}
\label{mu-def}
\mu = \frac{2\alpha}{1+\alpha^2},
\qquad
\alpha = (\pi \nu JS)^2
\end{equation}
(where $\nu$ is the DOS at the Fermi energy per one spin projection),
controls the strength of a single impurity ($\mu\to 0$ is the Born limit,
and $\mu \sim 1$ is the unitary limit) \cite{mu-comment},
while information about their concentration is contained in the parameter
\begin{equation}
\label{eta-def}
  \eta = \frac{\overline{n}_s \mu}{\pi\nu \Delta} ,
\end{equation}
where $\overline{n}_s$ is the average concentration of magnetic impurities.
In their original treatment, AG \cite{AG} considered the white-noise magnetic disorder
in the Born limit (many weak magnetic scatterers) that corresponds to
$\overline{n}_s\to\infty$ and $\mu\to 0$ at fixed $\eta$ (in this limit
$\eta=1/\Delta\tau_s$, with $\tau_s$ being the electron spin-flip time).
Equation (\ref{mu-def}) predicts a duality between the weak ($\alpha<1$)
and strong ($\alpha>1$) couplings: $\mu(\alpha)=\mu(1/\alpha)$.
However, the strong-coupling physics is more involved due to
partial screening of the spin of a magnetic impurity
by an unpaired quasiparticle leading to the formation
of a non-BCS ground state \cite{Sakurai,Balatsky-review}.
To avoid this complication, below we work in the weak coupling limit, $\alpha<1$.

In Eq.\ (\ref{eta-def}), $\Delta$ stands for the \emph{average value}\/ of the order parameter
in the presence of magnetic impurities. It is reduced compared to the magnetic-disorder-free case
and should be determined self-consistently. Randomness in locations of magnetic impurities
induces spatial fluctuations of $\Delta(\br)$ which will be discussed
in Sec.~\ref{SS:Delta-fluct} and Appendix~\ref{A:Delta}.

\begin{figure}
\includegraphics[scale=1]{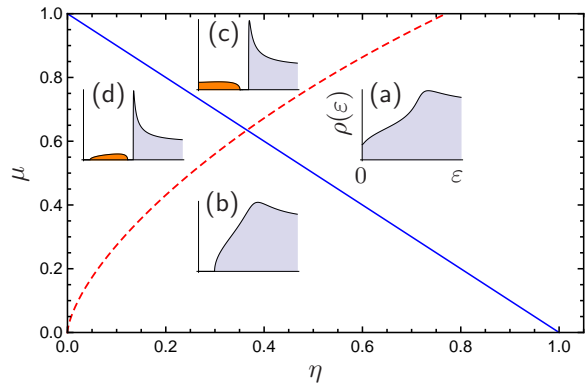}
\caption{
Regions of various behavior of the mean-field DOS $\rho(E)$ in the ($\eta$, $\mu$) plane.
The AGSR DOS at $E=0$ is zero (finite) below (above) the solid blue line.
The impurity band is resolved (merged with the continuum) above (below) the dashed red line.
Insets illustrate typical behavior of $\rho(E)$ in each region \cite{fig1etamu}. See Appendix~\ref{A:lines} for details.}
\label{F:etamudiag}
\end{figure}

\subsection{Mean-field theory}
\label{SS:AGSR}

The results of the mean-field calculation of the DOS by Abrikosov and Gor'kov \cite{AG}, Shiba \cite{Shiba}, and Rusinov \cite{Rusinov} (AGSR) can be summarized as follows.

\begin{figure}
\includegraphics[width=0.99\columnwidth]{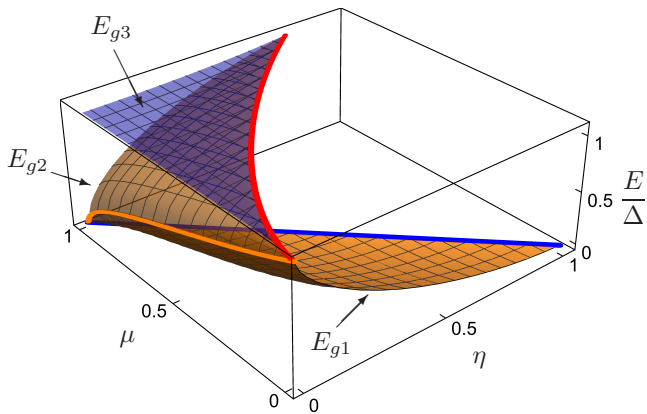}
\caption{
Position of the spectrum edges $E_{gi}$ (normalized to $\Delta$) vs.\ ($\eta$, $\mu$). Depending on the ($\eta$, $\mu$) values, the number of the spectrum edges varies from zero to three (the corresponding three sheets on the plot are denoted by $E_{g1}$, $E_{g2}$, and $E_{g3}$). The bold yellow line (lying in the $\eta=0$ plane) traces the position of the single-impurity localized state $E_0$ [Eq.\ (\ref{Ebound})] and joins the $E_{g1}$ and $E_{g2}$ sheets. The bold blue line lies in the $E=0$ plane (it is exactly the solid blue line in Fig.~\ref{F:etamudiag}) and marks the appearance of finite DOS at zero energy; here the $E_{g1}$ sheet terminates. The bold red line marks merging of the impurity band with the continuum (merging of the $E_{g2}$ and $E_{g3}$ sheets). This line starts from $E/\Delta=1$ at $\eta=\mu=0$, and rises slightly above the $E/\Delta=1$ level. Its projection on the ($\eta$, $\mu$) plane is exactly the dashed red line from Fig.~\ref{F:etamudiag}.}
\label{F:egs}
\end{figure}

Abrikosov and Gor'kov \cite{AG} considered suppression of the spectral gap $E_g$
(the lower edge for the continuum of Bogoliubov quasiparticles)
by weak magnetic impurities ($\mu\to 0$), finding
\begin{equation}
\label{EgAG}
  E_g^\text{AG} = \bigl( 1- \eta^{2/3} \bigr)^{3/2}\Delta .
\end{equation}
Later it was realized \cite{Yu,Soda,Shiba,Rusinov} that at a finite $\mu$, a single magnetic impurity
creates a subgap state with the energy
\begin{equation}
\label{Ebound}
  E_0 = \Delta \sqrt{\frac{1-\mu}{1+\mu}} = \Delta \frac{1-\alpha}{1+\alpha},
\end{equation}
localized at the length scale
\be
\label{L0}
  L_0 = \xi_0 (1-E_0^2/\Delta^2)^{-1/4} .
\ee
Equation (\ref{L0}) refers to diffusive superconductors \cite{com-L0},
with $\xi_0$ being the dirty-limit coherence length,
\begin{equation}
\label{xi-def}
\xi_0 = \sqrt{D/2\Delta} .
\end{equation}
Adding more impurities leads to the overlap of the states localized on different impurities,
and a well-defined impurity band between $E_{g1}$ and $E_{g2}$ is formed inside the superconducting gap.
The width of the band grows with increasing the impurity concentration (i.e., increasing $\eta$).
Shiba \cite{Shiba} and Rusinov \cite{Rusinov} described the properties
of the impurity band and showed that depending on the values of $\mu$ and $\eta$,
four possible scenarios indicated in Fig.~\ref{F:etamudiag} can be realized:
\begin{enumerate}
\item[(a)] no gap edges, gapless regime;
\item[(b)] AG regime with one spectrum edge $E_{g1}$ (the impurity band merged with the continuum);
\item[(c)] the impurity band touches zero, so the two spectrum edges
are the upper edge of the impurity band ($E_{g2}$) and the lower edge of the continuum ($E_{g3}$);
\item[(d)] the impurity band is detached both from zero and the continuum,
so there are two edges of the impurity band ($E_{g1}$ and $E_{g2}$) and the lower edge of the continuum ($E_{g3}$).
\end{enumerate}
Evolution of the gap edges, $E_{gi}(\eta,\mu)$,
demonstrating the transitions between the above regimes, is shown in Fig.~\ref{F:egs}.

Technically, in the mean-field theory, the DOS $\rho(E)$ (normalized to the normal-metallic value $\rho_0$) is given by
\begin{equation}
\label{rho-via-psi}
  \rho(E)/\rho_0 = \Im \sinh \psi ,
\end{equation}
where the energy-dependent spectral angle $\psi$ must be obtained from
the algebraic equation

\begin{equation} \label{F(psi)=0}
  F(\psi) = 0,
\end{equation}
with \cite{AG,Shiba,Rusinov,MS}
\begin{equation}
\label{F(psi)}
F(\psi) = - \frac E\Delta \cosh\psi + \sinh\psi -\frac\eta 2 \frac{\sinh 2\psi}{1-\mu \cosh 2\psi} .
\end{equation}
Since real $\psi$ leads to a vanishing DOS, the disappearance of a real solution
of Eq.\ (\ref{F(psi)=0}) marks a spectrum edge $E_g$ (either of the continuum
or of the impurity band). The equation for the determination of $E_g$
and positions of the lines separating the four regions in Fig.~\ref{F:etamudiag}
are discussed in Appendix \ref{A:lines}.

Note that the mean-field DOS structures similar to the ones presented in Fig.~\ref{F:etamudiag}, might also be realized in a diffusive superconductor with the order-parameter disorder, as shown recently in Ref.~\cite{Bespalov}. At the same time, Ref.~\cite{Bespalov} treated $\Delta(\mathbf r)$ as an external field without taking the self-consistency into account (similarly to the ROP model), while assuming small-scale inhomogeneities (with scale much smaller than the coherence length) and considering scattering on those order-parameter ``impurities'' in all orders of the perturbation theory ($T$-matrix approach). In our model, we take the self-consistency into account, and such pointlike order-parameter impurities are not realized.

\subsection{Results: subgap states}
\label{SS:results}

Hard mean-field gap edges are smeared by fluctuations,
leading to the average DOS sketched in Fig.~\ref{F:dos-smeared}. Provided that magnetic impurities are not too weak [see Eq.\ (\ref{cond*}) for the precise condition], the leading source of smearing at $E\to E_g$ is due to fluctuations of their concentration.
For larger $|E-E_g|$, this mechanism becomes less effective
and smearing due to mesoscopic fluctuations of potential disorder might dominate. In order to study the interplay of these mechanisms, we map the problem to the random order parameter model and calculate the average subgap DOS with the exponential accuracy as
\begin{equation}
\label{rho-S}
  \left< \rho(E) \right> \propto \exp( - \mathcal S_\text{inst} ),
\end{equation}
where $\mathcal S_\text{inst}$ is the action of the instanton with the broken replica symmetry.

Though the original problem is three-dimensional,
the effective dimensionality of the instanton, $d=0,1,2$, or 3, is determined from comparison of the sample dimensions to the instanton (optimal fluctuation) size $L_E$ given by Eq.\ (\ref{LE}) below.
In the present analysis, we restrict ourselves to the three- and zero-dimensional geometries. The cases $d=1$ and $d=2$ require special treatment due to the presence of multiple instanton solutions and will be reconsidered elsewhere \cite{STS}.

\subsubsection{Summary of the random order parameter model}

According to the general consideration of Ref.~\cite{SF}, subgap states in a wide class of disordered superconducting systems with a mean-field AG-like hard gap can be universally described by the random-order-parameter (ROP) model.
This scheme relies on the observation that at $E\to E_g$
any source of nonpotential disorder (random coupling constant \cite{LO_1971,MeyerSimons2001}, mesoscopic fluctuations of the order parameter \cite{FS}, infinitesimally weak magnetic impurities \cite{AG,LamacraftSimons}) effectively acts as a Gaussian random order-parameter field, $\Delta(\br) = \Delta + \Delta_1(\br)$, characterized by an appropriate correlation function
\be
\label{f-def}
  f(\bq) = \corr{\Delta_1\Delta_1}_\bq .
\ee
Once the mapping to the ROP model is identified, one can apply the known results \cite{SF} for the density of the subgap states, which are briefly reviewed below.

In the ROP model, a nonzero DOS in the gapped region originates from the interplay of potential disorder and disorder in $\Delta(\br)$, which is taken care of by the  parameter $K$. The instanton action $\mathcal S_\text{inst}$ is proportional [see Eq.\ (\ref{Sinst})] to $\mathcal S_0(K)$ given by
\begin{equation}
\label{S_0}
  \mathcal S_0(K)
  =
  \frac 16
  \int
  \left(
    \phi_2^3 - \phi_1^3
  \right)
  d^d \tilde{\mathbf r} ,
\end{equation}
where the functions $\phi_1(\tilde\br)$ and $\phi_2(\tilde\br)$ should be obtained from the system of coupled differential equations
\begin{subequations}
\label{SP12}
\begin{align}
-\tilde{\nabla}^2 \phi_1 +\phi_1-\phi_1^2 &= K(\epsilon)(\phi_2-\phi_1), \\
-\tilde{\nabla}^2 \phi_2 +\phi_2-\phi_2^2 &= K(\epsilon)(\phi_2-\phi_1) .
\end{align}
\end{subequations}
The system (\ref{SP12}) is characterized by a single dimensionless parameter
\be
\label{K-def}
  K(\epsilon) = \sqrt{\epsilon_*/\epsilon} ,
\ee
which is controlled by the dimensionless distance from the gap edge
\begin{equation}
\label{epsilon-def}
\epsilon = \frac{|E_g-E|}\Delta \ll 1
\end{equation}
(note that our definition of the dimensionless distance $\epsilon$ is different from $\varepsilon$ in Ref.~\cite{SF}, which was normalized by $E_g$; since here we consider the problem with several gap edges, we choose a more convenient normalization to $\Delta$).
The value of $\epsilon_*$ in Eq.\ (\ref{K-def}) is determined by the zero-momentum component of the correlation function of nonpotential disorder, $f(0)$ [see Eq.\ (\ref{epsilon*})].

In general, Eqs.\ (\ref{S_0})--(\ref{K-def}) determine the instanton action for any relation between $\epsilon$ and $\epsilon_*$.
However, for an arbitrary value of $K$, the system (\ref{SP12}) allows only for a numerical solution. Analytical treatment is possible in the limiting cases of large $K$ [close to the gap, $\epsilon\ll\epsilon_*$, see Eq.\ (\ref{subgap-LO})] and small $K$ [sufficiently far from the gap, $\epsilon\gg\epsilon_*$, see Eq.\ (\ref{subgap-MS})].
These limits refer to the situations when the subgap states are due to optimal fluctuations of either the nonpotential disorder (at $K\to\infty$) or potential disorder (at $K\to0$).
The instanton at an arbitrary $K$ describes an optimal fluctuation due to combined action of the potential and nonpotential disorder \cite{FS,SF}.

\subsubsection{General results for magnetic impurities}

As we demonstrate in Sec.~\ref{S:RSB}, the problem of Poissonian magnetic impurities also fits the phenomenology of the ROP model.
In order to apply the results of Ref.~\cite{SF}, one has
\begin{enumerate}
\item[(i)]
to generalize them to the case of an arbitrary function $F(\psi)$ [Eq.\ (\ref{F(psi)})] which determines the mean-field DOS (in Ref.~\cite{SF} only the AG case, $\mu=0$, was considered),
\item[(ii)]
to translate magnetic disorder to the language of the ROP model and to calculate the effective correlation function $f(\bq)$.
\end{enumerate}

Below we present the general expression for the subgap DOS, while the correlation function $f(\bq)$ in the case of magnetic impurities is discussed in Sec.~\ref{SSS:f}.
In the formulas below, the effective dimensionality of the instanton may take the values $d=3$ and $d=0$.

The instanton action
\begin{equation}
\label{Sinst}
  \mathcal S_\text{inst}
  =
  2
  g_\xi \frac{(2\epsilon \cosh\psi_g)^{(6-d)/4}}{|F''(\psi_g)|^{(2+d)/4}} \mathcal S_0(K)
\end{equation}
is proportional to $\mathcal S_0(K)$ given by Eq.\ (\ref{S_0}), with the prefactor depending on the curvature of the function $F(\psi)$ at $\psi_g$ corresponding to a gap edge $E_g$.
In Eq.\ (\ref{Sinst}), $g_\xi$ is the dimensionless (in units of $e^2/h$) conductance of the sample's section which is the hypercube of size $\xi_0$ in $d$ effective dimensions, while being limited by the sample size in the transverse directions. Denoting the volume of this section as $V_\xi = A_{3-d} \xi_0^d$, where $A_{3-d}$ is the ``cross section'' in $3-d$ reduced dimensions, we can write
\be
\label{gxi}
  g_\xi = 8\pi\nu \Delta V_\xi \gg 1.
\ee
Finally, the value of $\epsilon_*$ entering the definition of $K(\epsilon)$ in Eq.\ (\ref{K-def}) is given by
\begin{equation}
\label{epsilon*}
  \epsilon_*
  = \frac{g_\xi^2}{8}
  \left[ \frac{f(0)}{\Delta^2 V_\xi} \right]^2
  \frac{\sinh^4\psi_g}{|F''(\psi_g)| \cosh\psi_g}.
\end{equation}

Equations (\ref{rho-S}) and (\ref{Sinst}) provide a general description of the fluctuation DOS in the vicinity of the mean-field gap.
At $\epsilon\ll\epsilon_*$, the subgap states are due to optimal fluctuations of the concentration of magnetic impurities, and the result has a universal form \cite{SF,LO_1971}
\begin{equation}
\label{subgap-LO}
  \corr{\rho(E)}
  \propto
  \exp\left[ - \alpha_d(\eta,\mu) \,
  \frac{\Delta^2 V_\xi}{f(0)} \, \epsilon^{(8-d)/4} \right] .
\end{equation}
As $\epsilon$ grows, the role of mesoscopic fluctuations becomes increasingly important. In the regime of $\epsilon\gg\epsilon_*$ the subgap states are solely due to optimal fluctuations of potential disorder, with another universal behavior \cite{MS,SF}
\begin{equation}
\label{subgap-MS}
  \corr{\rho(E)}
  \propto
  \exp\left[ -\beta_d(\eta,\mu) \, g_\xi \, \epsilon^{(6-d)/4} \right] .
\end{equation}
The dimensionless functions $\alpha_d(\eta,\mu)$ and $\beta_d(\eta,\mu)$ are defined in Eqs.\ (\ref{alpha-d}) and (\ref{beta-d}), respectively.
Several limiting cases of Eqs.\ (\ref{subgap-LO}) and (\ref{subgap-MS}) will be discussed in Sec.~\ref{S:Limiting}.

\subsubsection{Parameter of the effective ROP model}
\label{SSS:f}

Reducing the problem of magnetic impurities to the ROP model valid in the vicinity of the mean-field gap is the most delicate issue. The reason is that there exist several physically distinct mechanisms which contribute to the effective correlation function (\ref{f-def}) characterizing the resulting ROP model.
One can distinguish three different contributions to $f(q)$ which enter additively (provided that the resulting gap smearing is relatively weak):
\begin{equation}
\label{f+f+f+f+f}
  f
  =
  \underbrace{f_{n_s}}_{\propto 1/g_\xi}
  \!\!
  +
  \underbrace{f_s + f_\text{MF}^\Delta}_{\propto 1/g_\xi^2}.
\end{equation}
The terms in this equation refer to the following types of effective inhomogeneities:
\begin{enumerate}
\item
$f_{n_s}(q)$ --- fluctuations due to inhomogeneity of the concentration of magnetic impurities;
\item
$f_s(q)$ --- fluctuations involving the triplet sector induced by a random spin orientation of magnetic impurities;
\item
$f_\text{MF}^\Delta(q)$ --- mesoscopic fluctuations of the order-parameter field $\Delta(\br)$ due to randomness in positions of potential impurities.
\end{enumerate}

The last two contributions, $f_s(q)$ and $f_\text{MF}^\Delta(q)$, were analyzed in the case of infinitesimally weak magnetic impurities ($\mu\to0$) in Ref.~\cite{SF}.
They are suppressed by the factor of $1/g_\xi^2$ typical for mesoscopic fluctuations \cite{UCF}
and thus are very small for a good metal.

Here, we focus on the term $f_{n_s}(q)$ which is proportional to $1/g_\xi$ but contains an additional small factor of $\mu$, vanishing in the Born limit ($\mu\to 0$). It gives the leading contribution to the correlation function (\ref{f+f+f+f+f}), provided magnetic impurities are not too weak:
\be
\label{cond*}
  \eta^{1/3}\mu \gg 1/g_\xi ,
\ee
which will be assumed thereafter.
Under this condition, the effective ROP correlation function $f(q)\approx f_{n_s}(q)$ is determined by fluctuations of the concentration of magnetic impurities, $n_s(\br)=\overline n_s+\delta n_s(\br)$.
In the zero-momentum limit, it can be written as
\be
\label{f=f+f}
  f(0) = \overline{n}_s [C(0) + C_\Delta(0)]^2 ,
\ee
where the first term in the sum is due to fluctuations of the spin-flip scattering rate (at a constant $\Delta$), whereas the second term is due to fluctuations of $\Delta(\br)$ (at a constant $\eta$). They are combined additively in Eq.\ (\ref{f=f+f}) since both are induced by $\delta n_s(\br)$.

The resulting expression for $C(0)$ is given by
\begin{equation}
\label{C-eta-0}
  C(0)
  =
  \left(\frac{\Delta^2 V_\xi}{\overline n_s} \frac{8\mu\eta}{g_\xi}\right)^{1/2}
  \frac{\cosh\psi_g}{1-\mu\cosh2\psi_g} ,
\end{equation}
where $\psi_g$ is the spectral angle at the gap edge [see Eq.\ (\ref{psig}) below].
The value of $C(0)$ depends on the particular gap edge considered, it is positive for $E_{g1}$ and negative for $E_{g2}$ and $E_{g3}$.
The implicit temperature dependence of $C(0)$ originates from that of $\psi_g$ since the parameter $\eta$ is expressed in terms of the temperature-dependent $\Delta(T)$ [Eq.\ (\ref{eta-def})].
The results by Silva and Ioffe \cite{Silva} correspond to the contribution to the DOS from $C(0)$ in the limit $\mu^{3/2}\ll \eta\ll 1$ (weak magnetic impurities in the AG regime), see Sec.~\ref{SS:Silva}.

The kernel $C_\Delta(q)$ describing the order-parameter fluctuations induced by $\delta n_s(\mathbf r)$ is calculated in Appendix \ref{A:Delta} [Eq.\ (\ref{Delta1nsfinal})]; it turns out to be positive.
 Since it involves the self-consistency condition, the result is temperature dependent.
In the limit of zero momentum and small temperatures,
$C_\Delta(0)$ is given by Eq.\ (\ref{C_Delta(0)-T0}).
In the gapped phase [regions (b) and (d) in Fig.~\ref{F:etamudiag}],
we find
\begin{equation}
\label{C-Delta-0}
  C_\Delta(0)
  =
  \left(\frac{\Delta^2 V_\xi}{\overline n_s} \frac{8\mu\eta}{g_\xi}\right)^{1/2}
  \frac{\pi}{2(1+\mu+\sqrt{1-\mu^2}) - \pi\eta}
  .
\end{equation}
Contrary to $C(0)$ [Eq.\ (\ref{C-eta-0})], the parameter $C_\Delta(0)$ describing the order-parameter fluctuations is the same for different gap edges $E_{gi}$.

The relative magnitude of the two contributions, $C(0)$ and $C_\Delta(0)$,
depends on $\eta$ and $\mu$, and on a particular gap edge considered.
Each of them can dominate in a certain region of parameters
as discussed in Sec.~\ref{S:Limiting}.

We emphasize here that our reduction to the effective ROP model [Eqs.\ (\ref{f=f+f})--(\ref{C-Delta-0})] holds irrespective of the resulting instanton dimensionality $d$ [$d$-dependent quantities $V_\xi$ and $g_\xi$ in the definitions of $C(0)$ and $C_\Delta(0)$ cancel each other].

In order to find the average subgap DOS at given values of $\eta$ and $\mu$ in the vicinity of a particular gap edge $E_g$,
one has to calculate first the mean-field gap angles $\psi_g$ [from Eq.\ (\ref{psig})],
the mean-field gaps $E_g$ [from Eq.\ (\ref{Eg})], and the derivative $F''(\psi_g)$ [from Eq.\ (\ref{F(psi)})].
These quantities determine the values of $f(0)$ [Eq.\ (\ref{f=f+f})],
$\alpha_d(\eta,\mu)$ [Eq.\ (\ref{alpha-d})],
and $\beta_d(\eta,\mu)$ [Eq.\ (\ref{beta-d})],
which govern the asymptotic behavior of the average DOS, Eqs.\ (\ref{subgap-LO}) and (\ref{subgap-MS}).
In some limiting cases, this procedure will be carried out in Sec.~\ref{S:Limiting}.
Finally, we remark that the above analysis was based on the mapping to the ROP model valid at $E\to E_g$. The condition of its validity will be discussed in Sec.\ \ref{SS:limits}.

\section{Field-theoretical approach}
\label{S:fieldtheory}

\subsection{Nonlinear sigma model}

In order to study the DOS in disordered superconductors with magnetic impurities,
we employ the standard sigma-model approach in the replica representation \cite{Finkelstein90,Belitz94},
applicable in the dirty limit ($\Delta\tau\ll 1$). It is formulated in terms of the
matrix field $Q$ describing the soft diffusive modes and the superconducting
order parameter field $\Delta$. The field $Q(\mathbf r)$ is a matrix in the tensor product
of the Matsubara-energies (E), replica (R), Nambu-Gor'kov (N), and spin (S) spaces.
The static (quantum fluctuations are neglected) order-parameter field $\Delta^a(\br)$ also carries a replica index.

The problem formulation is the same as in the paper by Marchetti and Simons \cite{MS}, and, similarly to Simons and co-workers \cite{MeyerSimons2001,LamacraftSimons,MS}, we employ the nonlinear sigma model technique (but its replica version instead of supersymmetric one).
At the same time, we treat the Poissonian averaging over magnetic impurities without any simplifications
(see Sec.\ \ref{SS:averaging}),
which is crucial for the correct determination of the replica-symmetry-breaking solutions.
The results of Ref.~\cite{MS} are then reproduced as a limiting case.
Another difference from Ref.~\cite{MS} is that we also take into account
order parameter inhomogeneities induced by magnetic impurities.
To address that self-consistently, we are forced to utilize the imaginary-time Matsubara
version of the sigma model, keeping the full energy space.

We consider the situation when the spin-flip scattering rate is much smaller than
the potential scattering rate, $\tau\ll\tau_s$; however, the strength of an individual
magnetic impurity (characterized by the parameter $\mu$) is not necessarily weak.
In this case, it is convenient to postpone averaging over magnetic disorder
to the final step of the derivation.
After averaging over potential disorder and integrating over fermions, the standard derivation \cite{Finkelstein90,Efetov,MS} leads to the expression for the partition function $\mathcal Z$,
\be
\mathcal Z = \int D\Delta \, DQ \, e^{-\mathcal S[\Delta,Q]},
\ee
written in terms of the imaginary-time action $\mathcal S$:
\begin{gather}
\label{SD}
\mathcal S
= \mathcal S_\Delta+ \frac{\pi\nu}{8\tau} \int d^3 \mathbf r \tr Q^2 - \frac 12 \Tr \ln G^{-1},
\\
  \mathcal S_\Delta
  =
  \frac{\nu}{\lambda T}
  \sum_{a=1}^n
  \int d^3 \br \,
  |\Delta^a(\br)|^2.
\end{gather}
Here, $\lambda$ is the Cooper-channel interaction constant,
$T$ is the temperature,
$n$ is the number of replicas,
$\tr$ stands for the trace over $\text{E}\otimes\text{R}\otimes\text{N}\otimes\text{S}$,
while $\Tr$ acts also in the coordinate space.

The inverse Green operator is given by
\begin{gather}
G^{-1} = G_0^{-1} + i \varepsilon \tau_3 + i \hat\Delta - J \mathbf S \boldsymbol\sigma  \tau_3,\\
G_0^{-1} = - \frac{p^2}{2m} + \mu_F + \frac i{2\tau} Q , \\
 \hat\Delta
  = \begin{pmatrix}
      0 & \Delta \\
      \Delta^* & 0 \\
    \end{pmatrix}_\text{N},
\end{gather}
where $\varepsilon$ is the Matsubara energy,
and $\tau_i$ and $\sigma_i$ are the Pauli matrices in the Nambu and spin spaces, respectively.
The matrix $Q$ is subject to the standard nonlinear constraint $Q^2 =1$, and obeys the symmetry \cite{Houzet_Skvortsov}
\be
\label{Q-constraint}
  Q = \tau_1 \sigma_2 Q^T \tau_1 \sigma_2 ,
\ee
where the transposition acts in the energy space as well.
The average DOS can be extracted from $\corr{Q_{\varepsilon\varepsilon}}$ analytically continued
to real energies $E$, see Eq.\ (\ref{rho-via-Q}) below.

The next step in the derivation of the sigma model is the expansion of the logarithm (justified by the diffusive limit). In the case of strong magnetic impurities, however, we must keep the magnetic part of the logarithm unexpanded \cite{MS}:
\begin{equation}
\label{ln}
\ln G^{-1} \approx \ln G_0^{-1} + i G_0 ( \varepsilon  \tau_3 + \hat\Delta ) + \ln (1-G_0 J \mathbf S \boldsymbol\sigma \tau_3).
\end{equation}
The resulting sigma-model action can be written as
\begin{equation}
\label{SSS}
  \mathcal S = \mathcal S_\Delta + \mathcal S_D + \mathcal S_\text{mag},
\end{equation}
where $\mathcal S_D$ is the standard diffusive action \cite{com-4},
\begin{equation}
\label{S0}
\mathcal S_D
= \frac{\pi\nu}8 \int d^3 \mathbf r \tr \bigl[ D (\nabla Q)^2 - 4 (\varepsilon \tau_3 + \Delta \tau_1) Q \bigr]
\end{equation}
(we choose $\Delta$ to be real),
and the magnetic part $\mathcal S_\text{mag}$ originates from the last term in Eq.\ (\ref{ln}) after averaging over magnetic impurities. It will be considered below.

\subsection{Averaging over magnetic disorder}
\label{SS:averaging}

If the distance between the magnetic impurities,
$\overline n_s^{-1/3}$,
is much larger than the mean free path due to potential (nonmagnetic) disorder $l$ \cite{l-com}, we can approximate
\begin{equation}
G_0(\mathbf r_i, \mathbf r_j) \approx -i\pi \nu \tau_3 Q(\mathbf r_i) \delta_{ij}.
\end{equation}
Then the magnetic part of the action becomes separable in the individual magnetic impurities \cite{MS}:
\begin{equation} \label{Ssep}
\mathcal S_\text{mag} \approx -\frac 12 \sum_i \tr\nolimits \ln \left( 1+ i\sqrt{\alpha} \, Q(\mathbf r_i)\tau_3 \boldsymbol\sigma \mathbf n_i \right) ,
\end{equation}
where the dimensionless parameter
$\alpha$ is defined in Eq.\ (\ref{mu-def}).
Replacing the full action $\mathcal S_\text{mag}$ by Eq.\ (\ref{Ssep})
is equivalent to the self-consistent $T$-matrix approximation \cite{AS} for the magnetic scattering, which treats all orders of scattering on a single impurity but neglects diagrams with intersecting impurity lines.

In what follows, we will neglect the effects of induced spin magnetization
and consider only the singlet sector of the theory, $Q=Q_0\sigma_0$
[such an approximation is justified under the condition (\ref{cond*})
when the leading source of the effective disorder is due to fluctuations
in the positions of magnetic impurities].
Then averaging over the direction of the impurity's magnetization $\mathbf{n}_i$
becomes trivial and we obtain
\begin{align}
\notag
  \mathcal S_\text{mag}
  =
  & - \frac 14 \sum_i \tr \ln \bigl[1 + \alpha Q(\mathbf r_i)\tau_3 Q(\mathbf r_i)\tau_3 \bigr]
\\
\label{Smag-via-ns}
  =
  & - \frac 14 \int d^3 \br \, n_s(\br)
  \tr \ln \bigl[1 + \alpha (Q\tau_3)^2 \bigr] ,
\end{align}
where we introduced the concentration of magnetic impurities [cf.\ Eq.\ (\ref{spin-density})]:
\be
\label{ns-def}
  n_s(\mathbf r) = \sum_i \delta(\mathbf r-\mathbf r_i) .
\ee

Performing Poisson averaging over magnetic disorder
with the help of the relation \cite{BGI}
\begin{equation}
\Bigl< \exp\Bigl\{ \sum_i f(\mathbf r_i) \Bigr\} \Bigr>
=
\exp \Bigl\{ \overline{n}_s \int d^3 \mathbf r \bigl[ e^{f(\mathbf r)}-1 \bigr] \Bigr\},
\end{equation}
where $\overline{n}_s$ is the average
concentration of magnetic impurities, we find the magnetic contribution to the sigma-model action:
\begin{equation}
  \mathcal S_\text{mag}^\text{av}
  =
  - \overline{n}_s \int d^3 \mathbf r
  \left( \exp
  \frac{\tr \ln [1+\alpha (Q \tau_3)^2]}4
  -1 \right) .
\label{Smagn0}
\end{equation}
An important feature of this expression
(where one can easily recognize the moment-generating
function for the Poisson distribution)
is its nonlinear dependence on $\tr\ln[1+\alpha (Q \tau_3)^2]$,
which will be crucial for the analysis
of subgap states.

Equation (\ref{Smagn0}) is the point where our derivation
starts to deviate from the one by Marchetti and Simons \cite{MS}.
Their approach is equivalent to replacing the exact action
$\mathcal S_\text{mag}^\text{av}$ by
Eq.\ (\ref{Smag-via-ns}),
where $n_s(\br)$ is substituted by its average value,
$n_s(\br)\mapsto\overline n_s$ [see Eq.\ (\ref{Smagn1}) below].
Such an approximation completely discards all effects due to fluctuations
of the concentration of magnetic impurities encoded
in higher powers of $\tr\ln[1+\alpha (Q \tau_3)^2]$.
This is justified only for replica-symmetric configurations of $Q$
(since each trace brings an additional power of $n$
which vanishes in the replica limit),
thus making it possible to reproduce the results of AGSR,
but is generally inapplicable for the analysis of the subgap states
associated with the replica-symmetry-breaking solutions.

In the problem of magnetic impurities, the field
\be
  \delta n_s(\br) = n_s(\br) - \overline n_s
\ee
can be identified as a \emph{primary fluctuator}\/
responsible for the formation of the subgap states through
the Larkin-Ovchinnikov mechanism \cite{LO_1971,SF}.
Its relevance for the problem of Poissonian magnetic impurities
was first recognized by Silva and Ioffe \cite{Silva}.

Another point which distinguishes our treatment from the analysis of Ref.~\cite{MS}
is the presence of the order-parameter field $\Delta(\br)$ that cannot be replaced
by its average value. Indeed, the field $\Delta(\br)$ adapts to inhomogeneity
of $n_s(\br)$, thus acting as an additional channel of disorder.
Its role will be analyzed below.

\subsection{Inhomogeneous order parameter and effective action}
\label{SS:Delta-fluct}

The action (\ref{SSS}) is a functional of the matter field
$Q(\br)$ and the order-parameter field $\Delta(\br)$.
Since the DOS is determined by $Q$,
our next task is to integrate out fluctuations of $\Delta(\br)$
and to derive an effective large-scale action $\mathcal S_\text{eff}[Q]$.
A routine approach would be to work
with the magnetic part $\mathcal S_\text{mag}^\text{av}$
[Eq.\ (\ref{Smagn0})] already averaged over disorder.
However, we find it more instructive to use $\mathcal S_\text{mag}$
in the initial form of Eq.\ (\ref{Smag-via-ns}) and perform the Poissonian
averaging after elimination of the order-parameter field.
This scheme clearly demonstrates that it is the field $\delta n_s(\br)$
which acts as a primary source of disorder, both directly
and via induced randomness in $\Delta(\br)$.

Due to self-consistency, the order parameter adapts to fluctuations
of the concentration of magnetic impurities. This can be described
in terms of a (replica-symmetric) linear response
of $\Delta_1(\br) = \Delta(\br) - \Delta$ to $\delta n_s(\br)$,
which in the momentum representation can be written as
\be
\label{Delta-ns}
  \Delta_1(\bq) = - C_\Delta(q) \delta n_s(\bq).
\ee
The temperature-dependent response kernel $C_\Delta(q)$ is calculated in Appendix \ref{A:Delta} [Eq.\ (\ref{Delta1nsfinal})] by summation over Matsubara energies.
The kernel $C_\Delta(q)$ is positive,
and the sign in Eq.\ (\ref{Delta-ns}) reflects that the order parameter
is suppressed in the regions where the concentration of magnetic impurities
exceeds its average value.
In real space, the kernel decays at the scale of the zero-temperature coherence length \cite{xi-com},
which is much smaller than the instanton size $L_E$ [Eq.\ (\ref{LE})] in the vicinity of the gap edge.
For this reason, only the zero-momentum limit of $C_\Delta(q)$ will be relevant below.
Then integrating out order-parameter fluctuations in the action (\ref{SSS})
produces the local term:
\be
\label{SOPF}
  \mathcal S_\text{OPF}
  =
  \gamma \int d^3 \br \, \delta n_s(\br) \tr \tau_1 Q ,
\ee
where
\be
\label{gamma-def}
  \gamma = \pi\nu C_\Delta(0)/2 .
\ee

Having eliminated fluctuations of the order-parameter field, we arrive at the action
$\mathcal S[Q] = \mathcal S_D + \mathcal S_\text{mag} + \mathcal S_\text{OPF}$
and are in a position to perform the final averaging over magnetic disorder.
Both $\mathcal S_\text{mag}$ and $\mathcal S_\text{OPF}$ are linear in $n_s$,
representing two ways inhomogeneities in the distribution of magnetic impurities
affect the system: through fluctuations of the overlap between the localized
states ($\mathcal S_\text{mag}$) and through the self-consistent modification
of $\Delta$ ($\mathcal S_\text{OPF}$).

Averaging $\mathcal S_\text{mag} + \mathcal S_\text{OPF}$
over the Poissonian distribution of $n_s(\br)$ is straightforward,
leading to the following term in the action
[which in the absence of the order-parameter fluctuations
reduces to $\mathcal S_\text{mag}^\text{av}$ given by Eq.\ (\ref{Smagn0})]:
\be
\label{Smag-final}
  \mathcal S_{n_s}
  =
  - \overline n_s \int d^3 \br
  \left[ e^{\tr \frac{\ln [1+\alpha (Q \tau_3)^2]}{4} - \gamma \tr \tau_1 Q}
  -1 + \gamma \tr \tau_1 Q \right] .
\ee
Equation (\ref{Smag-final}) should be considered as an effective action valid for $Q(\br)$ which changes slowly at the scale $\xi_0$. This condition guarantees the local relation between $Q(\br)$ and $\delta n_s(\br)$ as given by Eq.\ (\ref{SOPF}), which makes it possible to average the terms (\ref{SOPF}) and (\ref{Smag-via-ns}) on the same footing.
On the other hand, it should be understood that contrary to $n_s(\br)$, which is a set of $\delta$ functions,
$\Delta_1(\br)$ is a continuous function (a set of $\delta$ functions smeared at the scale $\xi_0$). For this reason, it is sufficient to keep the leading term in $\Delta$ in Eq.\ (\ref{ln}), as subleading terms are small in $\Delta\tau$ and may be neglected as usual.

Combined with the standard diffusive action $\mathcal S_D$
[Eq.\ (\ref{S0}) with a constant $\Delta$],
Eq.\ (\ref{Smag-final}) leads to the desired large-scale effective action for the field $Q$:
\be
\label{Seff[Q]}
  \mathcal S_\text{eff}[Q] = \mathcal S_D + \mathcal S_{n_s} .
\ee

\subsection{Simplification near the gap edge}

The nonlinear action $\mathcal S_{n_s}$ simplifies significantly in the vicinity of the spectrum edge, at $E\to E_g$, where the replica symmetry breaking (RSB) is weak, all traces in Eq.\ (\ref{Smag-final}) are small,
and the magnetic part of the action can be expanded in a power series:
\begin{equation}
  \mathcal S_{n_s}
  =
  \mathcal S_\text{mag}^{(1)}
  +
  \mathcal S_\text{mag}^{(2)}
  +
  \dots ,
\label{Smagn12}
\end{equation}
where
\begin{gather}
  \mathcal S_\text{mag}^{(1)}
  =
  - \frac{\overline{n}_s}{4} \int d^3 \mathbf r
  \tr \ln [1+\alpha (Q \tau_3)^2] ,
\label{Smagn1}
\\
  \mathcal S_\text{mag}^{(2)}
  =
  - \frac{\overline{n}_s}{2} \int d^3 \mathbf r
  \left\{
  \tr \left( \frac{\ln [1+\alpha (Q \tau_3)^2]}{4} - \gamma \tau_1 Q \right)
  \right\}^2 ,
\label{Smagn2}
\end{gather}
and omitted terms contain higher powers of the trace.
In Sec.~\ref{SS:universal}, we will see that the first two terms in the action (\ref{Smagn12})
are sufficient to describe the subgap tail states in the vicinity of the spectrum edge
(an analogous simplification has been recently carried out for the problem
of vacancies in chiral metals) \cite{vacancies}.
For larger deviations from the edge, the action $\mathcal S_{n_s}$
should be retained in its full form (\ref{Smag-final}).

The term $\mathcal S_\text{mag}^{(2)}$ can be naturally interpreted
as resulting from averaging of the action
$\mathcal S_\text{mag} + \mathcal S_\text{OPF}$
over Gaussian fluctuations of $\delta n_s$
specified by the correlation function
\begin{equation}
\label{corrnsns}
  \corr{\delta n_s(\br)\delta n_s(\br')} = \overline n_s \delta(\br-\br') .
\end{equation}
The fact that the Poissonian distribution of magnetic impurities
can be effectively described by Gaussian fluctuations should not be surprising.
In the vicinity of a spectrum edge, $E\to E_g$, the characteristic spatial scale
[$L_E$, see Eq.\ (\ref{LE}) below] diverges
and the corresponding instanton volume contains many magnetic impurities,
so that the central limit theorem applies.

Hence, for sufficiently small $\epsilon$ (the conditions are formulated in Sec.~\ref{SS:limits} below), the effective action (\ref{Seff[Q]})
can be approximated as
\begin{equation}
\label{Seff[Q]-approx}
  \mathcal S_\text{eff}[Q]
  \approx
  \mathcal S_D + \mathcal S_\text{mag}^{(1)} + \mathcal S_\text{mag}^{(2)}.
\end{equation}
Here, the first two terms are linear in the trace and lead to the AGSR theory
at the replica-symmetric saddle point (Sec.~\ref{SS:RSSP}),
whereas the last term is quadratic in the trace and
is responsible for gap fluctuations due to fluctuations in $n_s(\br)$.
The simplified action (\ref{Seff[Q]-approx}) will be used in Sec.~\ref{SS:universal} for the universal description of  subgap states near the gap edge.

\section{Instantons and subgap states}
\label{S:RSB}

\subsection{Replica-symmetric saddle point}
\label{SS:RSSP}

We start the analysis of the effective action (\ref{Seff[Q]-approx})
with the simplest replica-symmetric case.
The stationary replica-diagonal spin-singlet saddle point can be parametrized
in terms of the spectral angle $\theta_\varepsilon^a$ as
\begin{equation}
\label{Qparam}
Q_{\varepsilon\varepsilon'}^{ab} = \delta_{\varepsilon\varepsilon'} \delta_{ab} (\tau_3 \cos\theta_\varepsilon^a +\tau_1 \sin\theta_\varepsilon^a)\sigma_0 .
\end{equation}
Then only the linear-in-trace part of the action becomes important:
\begin{equation}
\label{S-linear}
\mathcal S_D + \mathcal S_\text{mag}^{(1)}
= \int d^3 \br \sum_{a=1}^n \sum_\varepsilon \mathcal{L}_a ,
\end{equation}
with the Lagrangian (written up to a constant term vanishing in the replica limit)
\begin{multline}
\label{L(theta)-def}
\mathcal{L}(\theta)
=
\frac{\pi \nu}2 \bigl[ D (\nabla \theta)^2 - 4(\varepsilon \cos\theta + \Delta \sin\theta)
\\
- (\Delta\eta/\mu) \ln(1+\mu \cos 2\theta) \bigr]  ,
\end{multline}
where the parameter $\mu$ is defined in Eq.\ (\ref{mu-def}).

The average DOS is calculated as
\begin{equation}
\label{rho-via-Q}
  \frac{\corr{\rho(E)}}{\rho_0}
  =
  \lim_{n\to0} \Re \frac{\tr_\text{R,N,S} \corr{Q_{EE}} \tau_3}{4n} ,
\end{equation}
where $\corr{Q_{EE}}$ is the expectation value of $Q$ with the action
$\mathcal S$, analytically continued to real energies: $\varepsilon\mapsto -iE$.
To simplify the analysis of the subgap states, it is convenient
to switch to a variable $\psi$ \cite{OSF01}:
\begin{equation} \label{thetapsi}
  \theta = \pi/2 +i\psi .
\end{equation}
In terms of $\psi_E^a$, the Lagrangian $\mathcal L$ acquires the form
\begin{multline}
\label{L(psi)-def}
\mathcal{L}(\psi)
=
- \frac{\pi \nu}2 \bigl[ D (\nabla \psi)^2 -4(E \sinh\psi - \Delta \cosh\psi)
\\{}
+ (\Delta\eta/\mu) \ln(1-\mu \cosh 2\psi) \bigr] .
\end{multline}

Varying the action (\ref{S-linear})
and searching for the replica-symmetric solution, we immediately
obtain an equation $F(\psi)=0$, where the function $F(\psi)$
is defined in Eq.\ (\ref{F(psi)}). Then Eq.\ (\ref{rho-via-Q}) reduces
to Eq.\ (\ref{rho-via-psi}), and we reproduce the results of AGSR discussed in Sec.~\ref{SS:AGSR}.
The action $\mathcal S_\text{mag}^{(2)}$ as well as higher-order terms
in Eq.\ (\ref{Smag-final}) do not affect the replica-symmetric solution.

\subsection{Universal description near the gap edge}
\label{SS:universal}

Subgap states are known to be associated with the RSB instantons \cite{MeyerSimons2001,SF}.
In the present case with many Matsubara energies involved,
we use the ansatz when the replica symmetry is violated
at a given energy $\varepsilon_0$. To respect the symmetry constraint (\ref{Q-constraint}) leading to $\psi_{\varepsilon}+\psi_{-\varepsilon}=0$, we have to include the energy $-\varepsilon_0$ as well:
\begin{equation}
\label{psi-RSB}
  \psi^a_\varepsilon(\br)
  =
  \begin{cases}
    \psi_1(\br), & \text{$\varepsilon=\varepsilon_0$ and $a=1$;} \\
    \psi_2(\br), & \text{$\varepsilon=\varepsilon_0$ and $a=2,\dots,n$;} \\
    - \psi_1(\br), & \text{$\varepsilon=-\varepsilon_0$ and $a=1$;} \\
    - \psi_2(\br), & \text{$\varepsilon=-\varepsilon_0$ and $a=2,\dots,n$;} \\
    \psi_\varepsilon(\br), & |\varepsilon|\neq\varepsilon_0.
  \end{cases}
\end{equation}
(A similar form of the RSB in the energy space
was considered in the context of energy level statistics
in random matrices \cite{KamenevMezard,KamenevAltland}.)
One can verify that such an ansatz is consistent with the saddle-point equations for the action (\ref{Seff[Q]}).
Note that although the replica symmetry
is assumed to be broken at the energies $\pm\varepsilon_0$, saddle-point solutions $\psi_\varepsilon(\br)$ at other energies acquire a nontrivial spatial dependence.
However, they do not influence
equations for $\psi_1(\br)$ and $\psi_2(\br)$ at $\varepsilon=\pm\varepsilon_0$, which are decoupled from
other energies. Performing analytic continuation, we can thus consider a single real energy $E=i\varepsilon_0$ \cite{SF}, and the role of energy $-\varepsilon_0$ would be to double the contribution
to the action in Eq.\ (\ref{S-linear-psi}) \cite{com-4}.

In order to get the instanton equation for the fields $\psi_1(\br)$ and $\psi_2(\br)$,
one has to substitute the ansatz (\ref{psi-RSB}) into the saddle-point equations for the action (\ref{Seff[Q]}).
The resulting system can be written as
\begin{multline}
\label{SP-mostgeneral}
  - \xi_0^2 \nabla^2 \psi_a - \frac{E}{\Delta} \cosh\psi_a
  + \left( 1 + \frac{2\eta\gamma}{\mu} \right) \sinh\psi_a
\\
  \pm \frac{\eta}{2\mu}
  \frac{\partial}{\partial\psi_a}
  \left[
  \sqrt{\frac{1-\mu\cosh2\psi_1}{1-\mu\cosh2\psi_2}}
  e^{- 4 \gamma (\cosh\psi_1-\cosh\psi_2)}
  \right]
  =
  0 ,
\end{multline}
where the positive (negative) sign in front of the last term
corresponds to $a=1$~(2).
The instanton equations (\ref{SP-mostgeneral})
are quite cumbersome and in a general situation (arbitrary $E$) can be treated only numerically.

Remarkably, the analysis simplifies considerably in the vicinity of the spectral gap,
$E\to E_g$
(the precise conditions will be formulated in Sec.~\ref{SS:limits} below),
where one can use the simplified action (\ref{Seff[Q]-approx})
and map the system onto the ROP model with a proper correlator $f(0)$.
The key point in this mapping is that near the gap edge the RSB is weak,
$\psi_1$ and $\psi_2$ are close to each other,
and the action (\ref{Seff[Q]-approx}) can thus be expanded in $\psi_1-\psi_2$.

In the replica limit ($n\to0$), the linear-in-trace part (\ref{S-linear}) becomes
\begin{equation}
\label{S-linear-psi}
  \mathcal S_D + \mathcal S_\text{mag}^{(1)}
  =
  2
  \int d^3 \br \bigl[ \mathcal{L}(\psi_1) - \mathcal{L}(\psi_2) \bigr] ,
\end{equation}
where the factor of 2 accounts for the doublet $\{\varepsilon_0,-\varepsilon_0\}$ in Eq.\ (\ref{psi-RSB}) \cite{com-4}.
Near the gap, the Lagrangians $\mathcal{L}(\psi_i)$ can be replaced by their Taylor
series near the mean-field solution $\psi_0$ [which satisfies $F(\psi_0)=0$]:
\begin{multline}
\label{Lpsii-Lpsi0}
  \mathcal{L}(\psi_i) - \mathcal{L}(\psi_0)
\approx
- \pi \nu \Delta \bigl[ \xi_0^2 (\nabla \psi_i)^2
\\{}
+ F'(\psi_0) (\psi_i-\psi_0)^2
+ F''(\psi_0) (\psi_i-\psi_0)^3/3
\bigr] .
\end{multline}
The cubic term ought to be retained since the coefficient in front of the quadratic term,
$F'(\psi_0) \approx F''(\psi_g) (\psi_0-\psi_g)$, vanishes at $E=E_g$:
\begin{equation}
\label{F'(psi0)}
  F'(\psi_0)
  \approx \sqrt{2 \frac{E-E_g}{\Delta} F''(\psi_g) \cosh\psi_g}
  \propto \epsilon^{1/2}
  ,
\end{equation}
where $\epsilon$ is defined in Eq.\ (\ref{epsilon-def}).
Depending on the values of $\eta$ and $\mu$, the mean-field spectrum can have up to three edges
(see Figs.~\ref{F:etamudiag} and~\ref{F:egs}, and Appendix~\ref{A:lines}),
with the mean-field gapped regions corresponding to $E<E_{g1}$, $E>E_{g2}$, and $E<E_{g3}$.
It can be shown that $F''(\psi_g)$ in all these cases has the same sign as $E-E_g$, so that the expression under the square root in Eq.\ (\ref{F'(psi0)}) is always positive.
Also with our accuracy we can replace $F''(\psi_0)$ by $F''(\psi_g)$ in Eq.\ (\ref{Lpsii-Lpsi0})
and replace the energy argument $E$ of this function by $E_g$ [so, in our formulas $F''(\psi_g)$ is always taken at $E=E_g$].

To complete the mapping to the ROP model,
consider the quadratic-in-trace part
$\mathcal S_\text{mag}^{(2)}$ in the action (\ref{Seff[Q]-approx}).
Though the $Q$ dependence of the two terms under the trace
in Eq.\ (\ref{Smagn2}) is different,
in the limit $E\to E_g$ both are proportional to $\psi_1-\psi_2$,
which allows us to write
\be
\label{S-quadratic-psi}
  \mathcal S_\text{mag}^{(2)}
  \approx
  - 8(\pi\nu)^2 f(0) \sinh^2\psi_g \int d^3 \br \, (\psi_1-\psi_2)^2 ,
\ee
where $f(0)$ acquires the form of Eq.\ (\ref{f=f+f})
with
\be
\label{C(0)}
  C(0)
  =
  \frac{\mu \cosh\psi_g}
  {(\pi\nu) (1-\mu\cosh2\psi_g)} .
\ee
The form of the prefactor in Eq.\ (\ref{S-quadratic-psi})
is chosen to emphasize that the same replica-mixing term
describes the ROP model (\ref{f-def}) near the gap edge.

To write the resulting action in the canonic form, we introduce the new fields
$\phi_{1,2}$ according to
\begin{equation}
\label{psi-phi}
  \psi_{1,2}(\mathbf r) = \psi_0 - \frac{2F'(\psi_0)}{F''(\psi_g)} \phi_{1,2}(\tilde{\mathbf r}),
\end{equation}
and switch to the dimensionless coordinate $\tilde{\mathbf r} = \mathbf r/L_E$.
Here, the length scale
\begin{equation}
\label{LE}
L_E = \frac{\xi_0}{\sqrt{F'(\psi_0)}} \propto \epsilon^{-1/4} ,
\end{equation}
where $F'(\psi_0)$ is given by Eq.\ (\ref{F'(psi0)}),
determines the instanton size which diverges at the gap edge.
In terms of the new variables, the action (\ref{Seff[Q]-approx}) acquires the form (\ref{Sinst}), with the dimensionless action
\begin{multline}
\label{S_0-full}
  \mathcal S_0(K) =
  \int d^d\tilde\br
  \biggl[
    \frac{(\tilde\nabla \phi_2)^2}{2} + \frac{\phi_2^2}{2} - \frac{\phi_2^3}{3}
\\{}
  - \frac{(\tilde\nabla \phi_1)^2}{2} - \frac{\phi_1^2}{2} + \frac{\phi_1^3}{3}
  - K \frac{(\phi_1-\phi_2)^2}{2}
  \biggr]
  .
\end{multline}
The strength of replica mixing is controlled by the dimensionless parameter
\begin{equation}
  K(\epsilon)
  =
  \frac{4\pi\nu f(0) \sinh^2\psi_g} {\Delta F'(\psi_0)}
  =
  \sqrt{\frac{\epsilon_*}\epsilon} ,
\end{equation}
where the crossover scale $\epsilon_*$
can be represented in the form (\ref{epsilon*}).

Varying the action (\ref{S_0-full}), we arrive at the universal system (\ref{SP12}) of coupled differential equations for $\phi_{1,2}(\tilde\br)$. At the solution, the action $\mathcal S_0(K)$ can be written in the compact form of Eq.\ (\ref{S_0}).

\subsection{Instanton action versus $K$}
\label{SS:instantonvsK}

Below, we briefly overview the properties of the system (\ref{SP12}) obtained in Refs.\ \cite{FS,SF}.
We consider only the $d=3$ and $d=0$ cases, while the ROP model with $d=1$ and $d=2$ will be studied elsewhere \cite{STS}. Equations~(\ref{SP12}) can be easily analyzed in the limits of small and large $K$,
where analytic expressions for $\mathcal S_0(K)$ are possible
[Eqs.\ (\ref{s-d}) and (\ref{SK-LO}), respectively].
For intermediate values of $K$, the system should be solved numerically,
with the instanton action gradually
interpolating between the limiting values.
Following Refs.~\cite{LO_1971,LamacraftSimons,MS,Silva}, for $d=3$
we consider only spherically-symmetric instanton solutions.
The existence of a less symmetric instanton
with a smaller action cannot be excluded a priori
and requires a separate investigation.

\subsubsection{Zero-dimensional geometry}
\label{SSS:0D}

\begin{figure}[t]
\includegraphics[width=\columnwidth]{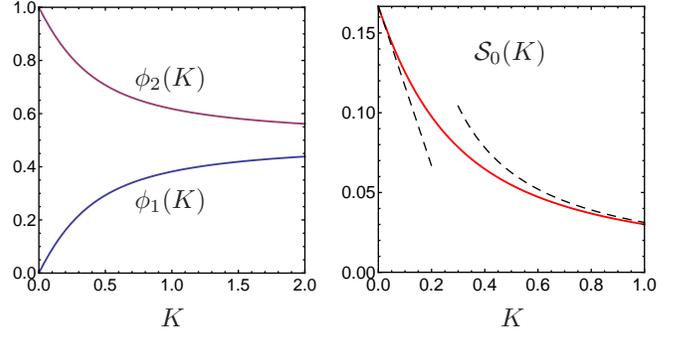}
\caption{
Instanton in the small-grain geometry.
Left panel: solutions $\phi_1$ (lower curve) and $\phi_2$ (upper  curve)
of the system (\ref{SP12}) vs.\ the parameter $K$.
Right panel: $K$ dependence of the action ${\mathcal S_0}(K)$
[dashed lines show the asymptotic behavior, see Eq.\ (\ref{S0D-asymp})].}
\label{F:0D}
\end{figure}

We start the analysis of the $K$-dependence of the instanton action
with the simplest zero-dimensional case realized for superconducting
grains smaller than the instanton size, $L_E$.
In this case, $\left< \rho(E) \right> \propto\exp(-{\mathcal S}_\text{inst})$
gives the DOS averaged over an ensemble of grains \cite{Vavilov,Silva}.
Neglecting the gradient terms in Eqs.\ (\ref{SP12}),
we arrive at a system of algebraic equations which can be easily solved:
\be
  \phi_{1,2} = 1/2 \pm K \mp \sqrt{K^2+1/4} ,
\ee
where the signs are chosen in order to provide a positive action.
Calculating the instanton action
with the help of Eq.\ (\ref{S_0}), we obtain
\be
  {\mathcal S_0}(K)
  =
  \frac{(4K^2+1)^{3/2}-K(8K^2+3)}{6} .
\ee
Evolution of the solutions $\phi_{1,2}$ and the action ${\mathcal S_0}$
with the parameter $K$ is shown in Fig.~\ref{F:0D}.
The asymptotic behavior,
\be
\label{S0D-asymp}
  {\mathcal S_0}(K)
  =
  \begin{cases}
    1/6 - K/2 + \dots, & K\ll1 ; \\
    1/32K + \dots, & K\gg1 ;
  \end{cases}
\ee
is depicted by the dashed lines in Fig.~\ref{F:0D}(b).

\subsubsection{Instanton in the limit $K\to0$}
\label{SSS:K-small}

In the limit $K\to0$ (i.e., far enough from the gap edge, $\epsilon \gg \epsilon_*$), Eqs.\ (\ref{SP12}) decouple,
yielding a single equation
\begin{equation}
\label{Usadel0}
  - \tilde\nabla^2 \phi + \phi - \phi^2 = 0
\end{equation}
both for $\phi_1(\tilde\br)$ and $\phi_2(\tilde\br)$.
This equation has a bounce solution $\varphi_\text{inst}^{(d)}(\tilde r)$
vanishing for $\tilde r\to\infty$ (in 1D, this solution is known explicitly, while for other dimensions it can be found numerically).
The action (\ref{S_0}) is minimized by taking the trivial
solution $\phi_1(\tilde\br)=0$ for the first replica and the
bounce solution, $\phi_2(\tilde\br) = \vp_\text{inst}^{(d)}(\tilde r)$,
for other replicas. The dimensionless instanton action $\mathcal S_0(0)=s_d$
is then given by the number
[the value of $s_0$ inferred from Eq.\ (\ref{S0D-asymp}) is added here for completeness]
\begin{equation}
\label{s-d}
  s_d
  \equiv
  \frac 16
  \int \bigl[ \vp_\text{inst}^{(d)}(\tilde r) \bigr]^3
  d^d \tilde\br
  =
  \begin{cases}
    1/6, & d=0 , \\
    43.7, & d=3 .
  \end{cases}
\end{equation}
Finally, with the help of Eq.\ (\ref{Sinst}),
we obtain Eq.\ (\ref{subgap-MS}) for the DOS, where
\begin{equation}
\label{beta-d}
  \beta_d(\eta,\mu)
  =
  2
  s_d \frac{(2\cosh\psi_g)^{(6-d)/4}}{|F''(\psi_g)|^{(2+d)/4}} .
\end{equation}

\subsubsection{Instanton in the limit $K\to\infty$}
\label{SSS:K-large}

In the limit $K\to\infty$ (i.e., close to the gap edge, $\epsilon \ll \epsilon_*$),
the RSB is weak: $\phi_1-\phi_2\to 0$. Due to a remarkable dimensional reduction \cite{Silva,SF},
a nontrivial optimal fluctuation for $\phi(\tilde\br)=\phi_1(\tilde\br)\approx\phi_2(\tilde\br)$
in $d$ dimensions is just the bounce solution of Eq.\ (\ref{Usadel0})
in $d-2$ dimensions \cite{com-d-2}:
$\phi^{(d)}(\tilde r)=\vp_\text{inst}^{(d-2)}(\tilde r)$.
The instanton action (\ref{S_0}) in the limit $K\to\infty$ then reads
\begin{equation}
\label{SK-LO}
  \mathcal S_0(K)
  = c_d/K ,
\end{equation}
where $c_d$ is the dimensionality-dependent constant \cite{SF}
[the value of $c_0$ inferred from Eq.\ (\ref{S0D-asymp}) is added here for completeness]:
\begin{equation}
\label{c-d}
  c_d
  = 2 \int
  \biggl(\frac{\partial\vp^{(d-2)}_\text{inst}(\tilde r)}{\partial \tilde r}\biggr)^2
  \frac{d^d \tilde \br}{\tilde r^2}
  =
  \begin{cases}
    1/32, & d=0 , \\
    24\pi/5, & d=3 .
  \end{cases}
\end{equation}
Substituting Eq.\ (\ref{SK-LO}) into the action (\ref{Sinst}),
we arrive at Eq.\ (\ref{subgap-LO}) with
\begin{equation}
\label{alpha-d}
  \alpha_d(\eta,\mu)
  =
  4 c_d \frac{(2 \cosh\psi_g)^{(8-d)/4}}{|F''(\psi_g)|^{d/4} \sinh^2\psi_g} .
\end{equation}

\subsection{Limits of the universal description}
\label{SS:limits}

Here, we summarize conditions on $\epsilon$ which allow us to derive the universal description near the gap edge developed in Sec.~\ref{SS:universal}.

This description is based on
(i)~an expansion of the linear-in-trace action, defined by Eqs.\ (\ref{S-linear}) and (\ref{L(psi)-def}), in powers of $(\psi_i-\psi_0)$, and accompanying approximations leading to the universal form, given by Eqs.\ (\ref{S-linear-psi}) and (\ref{Lpsii-Lpsi0}),
(ii)~expansion of the quadratic-in-trace action (\ref{Smagn2}), leading to the universal form of the replica-mixing term (\ref{S-quadratic-psi}),
(iii)~simplification of the full
action (\ref{Seff[Q]}) to the form (\ref{Seff[Q]-approx}), with only linear and quadratic in trace
terms being retained.

Considering approximation (i), we immediately see that the expansion in powers of $(\psi_i-\psi_0)$ requires
\be
\label{epsilon1}
  \epsilon\ll \epsilon_1,
  \qquad
  \epsilon_1 \sim
  \frac{|F''(\psi_g)|}{\cosh\psi_g} ,
\ee
with the characteristic value of $\epsilon_1$ estimated from Eqs.\ (\ref{F'(psi0)}) and (\ref{psi-phi}).
At the same time, one can check that approximation (\ref{F'(psi0)}) itself requires
\begin{equation} \label{epsilon2}
\epsilon\ll \epsilon_2,
\qquad
\epsilon_2 \sim \frac{\epsilon_1}{\left[ \tanh\psi_g +F'''(\psi_g)/F''(\psi_g) \right]^2}.
\end{equation}
Actually, it can be shown that conditions (\ref{epsilon1}) and (\ref{epsilon2}) also justify other approximations related to the linear-in-trace action [neglecting the fourth-order $(\psi_i-\psi_0)^4$ term in Eq.\ (\ref{Lpsii-Lpsi0}) and replacing $F''(\psi_0)$ at energy $E$ by $F''(\psi_g)$ at energy $E_g$]. For $\epsilon\gtrsim\epsilon_1,\epsilon_2$, the difference $(\psi_i-\psi_0)$ becomes
larger than 1 and the hyperbolic functions in Eq.\ (\ref{L(psi)-def}) should be retained in their full form.

With parametrization (\ref{Qparam}) and (\ref{thetapsi}), the quadratic-in-trace action (\ref{Smagn2}) at a single real energy $E$ takes the form
\begin{equation}
\mathcal S_\mathrm{mag}^{(2)} = -\frac{\overline{n}_s}2 \int d^3 \mathbf r \left[ \tr_\mathrm{R} \Phi(\psi) \right]^2,
\end{equation}
with
\begin{equation}
\Phi(\psi) = \frac 12 \ln(1-\mu\cosh 2\psi) - 4\gamma \cosh\psi.
\end{equation}
With this notation, approximation (ii) requiring that the cubic $(\psi_1-\psi_2)^3$ term can be neglected in Eq.\ (\ref{S-quadratic-psi}), imposes an additional requirement
\begin{equation} \label{epsilon3}
\epsilon\ll \epsilon_3,
\qquad
\epsilon_3 \sim \epsilon_1 \left[ \frac{\Phi'(\psi_g)}{\Phi''(\psi_g} \right]^2.
\end{equation}

Finally, approximation (iii) implying that $\mathcal S_\mathrm{mag}^{(3)}$ can be neglected, requires
\begin{equation} \label{epsilon4}
\epsilon\ll \epsilon_4,
\qquad
\epsilon_4 \sim \epsilon_1 \left[ \frac 1{\Phi'(\psi_g)} \right]^2.
\end{equation}

Thus, the conditions (\ref{epsilon1}), (\ref{epsilon2}), (\ref{epsilon3}), and (\ref{epsilon4}) determine the upper limit of applicability for our universal description near the gap edge. On the other hand, the lower limit is set by the condition $\mathcal S_\mathrm{inst} \gg 1$ ensuring validity of the saddle-point approximation (this condition is violated in the fluctuation region in very close vicinity of the mean-field gap edge).

\begin{figure*}[t]
\centering{\includegraphics[width=0.99\textwidth]{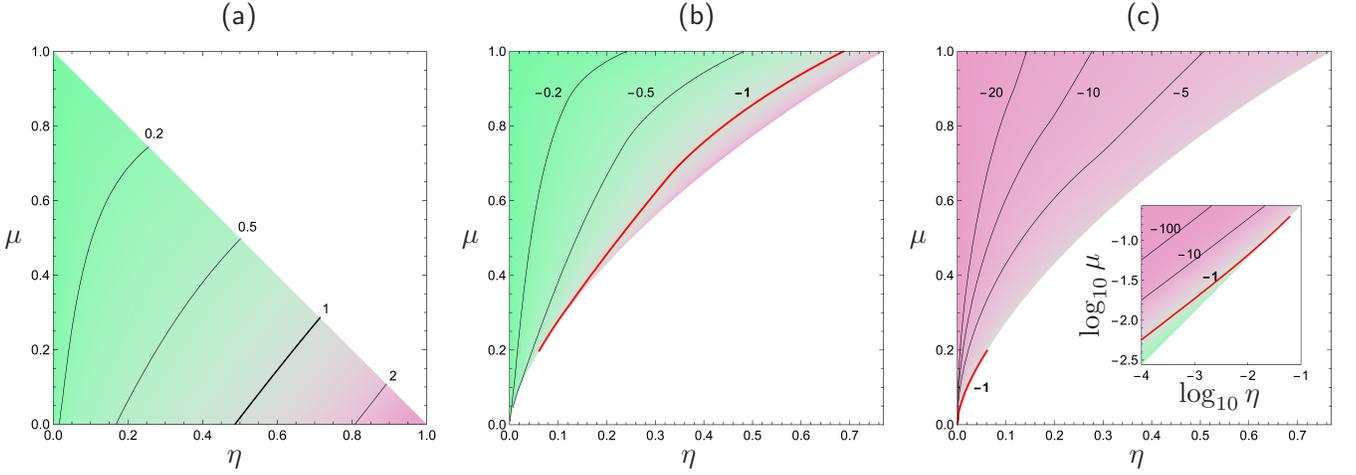}}
\caption{
The ratio $C_\Delta(0)/C(0)$ for the spectrum edges
(a) $E_{g1}$,
(b) $E_{g2}$, and
(c) $E_{g3}$
as a function of $\eta$ and $\mu$ at zero temperature.
The inset in panel (c) shows the region of small $\eta$ and $\mu$
in logarithmic scale.
Regions where gap smearing is mainly due to fluctuations
of the spin-flip scattering rate (of the order parameter)
are shown by green (pink). The red line in panels (b) and (c)
corresponds to $C(0)+C_\Delta(0)=0$ when both effects exactly
compensate each other and $f(0)=0$. In this case, gap smearing is
completely determined by mesoscopic fluctuations of potential disorder.
}
\label{F:f/f-Eg}
\end{figure*}

\section{Subgap states: Limiting cases}
\label{S:Limiting}

The results for the instanton action obtained above are valid for arbitrary $\mu$ (strength of individual magnetic impurities) and $\eta$ (their concentration), and apply in the vicinity of any of the three possible spectrum edges.
The general recipe for calculating the instanton action
is outlined at the end of Sec.~\ref{SS:results}.

In the Larkin-Ovchinnikov regime, at $\epsilon\ll\epsilon_*$, the subgap DOS is determined by optimal fluctuations of the concentration of magnetic impurities characterized by
the parameter $f(0)=\overline n_s [C(0)+C_\Delta(0)]^2$.
The two terms here correspond to fluctuations of the pair-breaking rate ($\eta$) and the order parameter ($\Delta$),
both being induced by $\delta n_s(\br)$.
The ratio of these two contributions, $C_\Delta(0)/C(0)$,
calculated numerically is presented in Fig.~\ref{F:f/f-Eg}.
It is positive for $E_{g1}$ and,
quite surprisingly,
negative for $E_{g2}$ and $E_{g3}$, indicating that in the latter cases fluctuations in $\eta$ and $\Delta$ partially compensate each other.

Different signs of $C(0)$ and $C_\Delta(0)$ for different gap edges
can be qualitatively understood by analyzing the mean-field expressions for $E_g$ available in the limiting cases: while $E_g$ is generally an increasing function of $\Delta$,
it decays with $\eta$ for $E_{g1}$ [Eqs.\ (\ref{EgAG}) and (\ref{Eg1g2})]
and grows with $\eta$ for $E_{g2}$ and $E_{g3}$ [Eqs.\ (\ref{Eg1g2}) and (\ref{Eg3})].
Since a local increase of $\delta n_s(\br)$ suppresses $\Delta$ and enhances $\eta$, it leads to the decrease of $E_{g1}$, whereas its effect on $E_{g2}$ and $E_{g3}$ is determined by the competition of two terms with opposite signs.
The two effects completely compensate each other, $C(0)+C_\Delta(0)=0$, at the red curve in Figs.~\ref{F:f/f-Eg}(b) and \ref{F:f/f-Eg}(c).
In the language of optimal fluctuations, subgap states originate from fluctuations of $\delta n_s(\br)$ which locally shift $E_g$ to the classically forbidden region.
From the above analysis, it follows that the optimal fluctuation in $\delta n_s(\br)$ is positive except for narrow regions below the red curve in Figs.~\ref{F:f/f-Eg}(b) and \ref{F:f/f-Eg}(c).

Further analytical progress is possible in certain limiting cases discussed below.

Sufficiently far from the gap edge, at $\epsilon\gg\epsilon_*$
(while still $\epsilon \ll \epsilon_1,\epsilon_2,\epsilon_3,\epsilon_4$ so that our approximations are valid, see Sec.~\ref{SS:limits}),
the average DOS is given by Eq.\ (\ref{subgap-MS}).
With the expression (\ref{beta-d}) for $\beta_d(\eta,\mu)$,
we reproduce the results by Marchetti and Simons \cite{MS,MScomment}.
This is only a far asymptotics for the DOS due to very rare fluctuations
of the potential disorder, while the main part of the subgap DOS is determined
by the Larkin-Ovchinnikov-like contribution (\ref{subgap-LO}) due to
fluctuations in the concentration of magnetic impurities.

The effective instanton dimensionality below takes the values $d=3$ and $d=0$.

\subsection{Weak magnetic impurities (small $\mu$)}
\label{SS:Silva}

Consider now the gapped AG regime
(magnetic impurities are weak, almost in the Born limit, so that there is only one spectrum edge, $E_{g1}$):
\begin{equation} \label{musmall}
\mu \ll \eta^{2/3} < 1 ,
\end{equation}
corresponding to the bottom of the region (b) in Fig.~\ref{F:etamudiag}.
In this limit, Eqs.\ (\ref{psig}) and (\ref{Eg}) simplify
and we obtain the standard AG solution, $\cosh\psi_g \approx 1 / \eta^{1/3}$,
with the gap $E_g^\text{AG}$ given by Eq.\ (\ref{EgAG})
and $F''(\psi_g) \approx -3 \eta^{1/3} (1-\eta^{2/3})^{1/2}$.

The parameter of the effective ROP model is given by
Eq.\ (\ref{f=f+f}), where $C_\Delta(0)$ is temperature-dependent. To simplify the analysis, we consider the $T=0$ case.
Then using Eqs.\ (\ref{C-eta-0}) and (\ref{C-Delta-0}), we obtain
\be
\label{DOS_LOeta}
  \frac{f(0)}{\Delta^2 V_\xi}
  \approx
  \frac{8\mu\eta^{1/3}}{g_\xi}
  \left(
    1
    +
    \frac{\pi\eta^{1/3}}{4-\pi\eta}
  \right)^2
  ,
\ee
where the two terms in the brackets correspond to the
contribution of $C(0)$ and $C_\Delta(0)$,
respectively.
In the limit of weak pair-breaking ($\eta\ll 1$), the value of $f(0)$ is determined mainly
by $C(0)$, which describes fluctuations of the overlap of the states localized at different
magnetic impurities due to fluctuation in their concentration (fluctuations of $\eta$).
On the other hand, near the gap closing, at $\eta\sim 1$,
the order-parameter fluctuations induced by magnetic impurities
give a comparable contribution
[with a moderately large numerical factor $C_\Delta(0)/C(0)\approx 3.6$ as $\eta\to1$].
The ratio $C_\Delta(0)/C(0)$ for the spectrum edge $E_{g1}$
is shown in Fig.~\ref{F:f/f-Eg}(a).
With the help of Eqs.\ (\ref{epsilon*}) and (\ref{DOS_LOeta}),
we obtain for the crossover energy:
\begin{equation}
  \epsilon_*
  =
  \frac{8\mu^2 (1-\eta^{2/3})^{3/2}}{3\eta^{2/3}}
  \left(
    1
    +
    \frac{\pi\eta^{1/3}}{4-\pi\eta}
  \right)^4
 .
\end{equation}

The main asymptotics of the DOS tail at $\epsilon\ll \epsilon_*$ is governed
by the action
\begin{equation} \label{DOS_LO}
\mathcal S_\mathrm{inst} = \frac{16 c_d}{6^{d/4} (1-\eta^{2/3})^{1+d/8}} \frac{\Delta^2 V_\xi}{f(0)} \epsilon^{(8-d)/4},
\end{equation}
with $f(0)$ given by Eq.\ (\ref{DOS_LOeta}).
In the case of weak spin-flip scattering, $\eta\ll 1$,
Eq.\ (\ref{DOS_LO}) simplifies to
\begin{equation}
\label{Sinst-LO}
  \mathcal S_\text{inst}
  =
  \frac{c_d}\mu \frac{2\cdot 6^{-d/4}}{\eta^{1/3}} g_\xi \epsilon^{(8-d)/4}
  =
  \tilde a_d \frac{\overline n_s V_\xi}{\eta^{4/3}} \epsilon^{(8-d)/4} ,
\end{equation}
where $\tilde a_d = 16\cdot 6^{-d/4} c_d$.
Equation (\ref{Sinst-LO}) coincides (within a few percent accuracy, probably due to numeric uncertainty
in the determination of the instanton action)
with the result of Silva and Ioffe \cite{Silva},
who considered the optimal fluctuation of the concentration of magnetic impurities.

The far asymptotics of the DOS tail at $\epsilon\gg \epsilon_*$ is determined by
the instanton action
\begin{equation} \label{DOS_MS}
\mathcal S_\mathrm{inst} = s_d g_\xi \frac{4\cdot 6^{(2-d)/4}}{3\eta^{2/3} (1-\eta^{2/3})^{(2+d)/8}} \epsilon^{(6-d)/4},
\end{equation}
which exactly coincides with the result of Lamacraft and Simons \cite{LamacraftSimons}.

Analyzing the upper-bound applicability conditions for $\epsilon$, formulated in Sec.~\ref{SS:limits}, we find that the most restrictive one is $\epsilon\ll \epsilon_2$, while
\begin{equation}
\frac{\epsilon_*}{\epsilon_2} \sim \left( \frac\mu{\eta^{2/3}} \right)^2 \ll 1.
\end{equation}
This means that our theory based on the universal description (see Sec.~\ref{SS:universal}) can trace both the main asymptotics of the tail due to magnetic disorder (at $\epsilon< \epsilon_*$) and its far asymptotics due to potential disorder (at $\epsilon> \epsilon_*$).

The physical meaning of $\epsilon_2$ becomes transparent in the case of weak spin-flip scattering, $\eta\ll 1$. In this limit, $\epsilon_2 \sim \eta^{2/3}$, which is of the same order as the mean-field smearing of the gap edge.

\subsection{Small concentration limit (small $\eta$)}

Here, we consider the limit of small impurity concentration,
\begin{equation} \label{etasmall}
\eta^{2/3}\ll\mu,
\end{equation}
corresponding to the left border of the region (d) in the phase diagram of Fig.~\ref{F:etamudiag}.
In this regime, a narrow impurity band is formed, and we study fluctuation smearing
of its edges at $E\to E_{g1}-0$ and $E\to E_{g2}+0$, as well as
smearing of the continuum hard-gap edge at $E\to E_{g3}-0$.

\subsubsection{Smearing of the impurity band
($E_{g1}$ and $E_{g2}$)}
\label{SS:band}

In the limit (\ref{etasmall}), the values of the spectral angles determining the mean-field edges of the impurity band are given by
\be
  \psi_{g1,g2}
  \approx
  \frac12 \arccosh \frac 1\mu
  \mp
  \frac{(1+\mu)^{1/4}}{2^{5/4}\mu^{3/4}} \sqrt\eta.
\ee
The edges of the impurity band are then expressed as
\be
\label{Eg1g2}
  \frac{E_{g1,g2}}\Delta
  \approx
  \frac{E_0}\Delta \mp
    \frac{2^{3/4}\mu^{1/4}}{(1+\mu)^{3/4}} \sqrt\eta,
\ee
where $E_0$ is the energy of the single-impurity bound state, Eq.\ (\ref{Ebound}).
From Eq.\ (\ref{F(psi)}), we find
\be
  F''(\psi_{g1,g2}) \approx \mp \frac{2^{11/4}\mu^{5/4}}{(1+\mu)^{3/4}} \frac{1}{\sqrt\eta}.
\ee
Smearing of the impurity band
determined by the action (\ref{Sinst})
is thus symmetric in the main order.

As it is shown in Figs.~\ref{F:f/f-Eg}(a) and \ref{F:f/f-Eg}(b) [see also Eqs.\ (\ref{C-eta-0}) and (\ref{C-Delta-0})],
$C_\Delta(0)/C(0)\ll1$
for small $\eta$.
Thus the contribution of $C_\Delta(0)$ to $f(0)$
can be neglected at zero temperature [the exact condition $\eta\ll\mu^{1/2}$ is certainly satisfied in the limit (\ref{etasmall})], and we find
\be
  \frac{f(0)}{\Delta^2 V_\xi}
  \approx
  \frac{2^{5/2} \mu^{3/2}}{g_\xi(1+\mu)^{1/2}(1-\mu)}
\ee
independently of the value of $\eta$.
For the crossover energy, we obtain
\begin{equation}
\epsilon_* = \frac{\mu^{1/4} \sqrt\eta}{2^{9/4} (1+\mu)^{3/4}}.
\end{equation}

Analyzing the upper-bound applicability conditions for $\epsilon$, formulated in Sec.~\ref{SS:limits}, we find that it is sufficient to require $\epsilon\ll \epsilon_2$, while $\epsilon_2 \sim \epsilon_*$.
This means that our results based on the universal description (see Sec.~\ref{SS:universal}) are valid only in the regime of the main asymptotics, $\epsilon< \epsilon_*$.
This asymptotics of the DOS
[Eq.\ (\ref{subgap-LO})] is governed by the action
\begin{equation}
\label{subgap-band-LO}
  \mathcal S_\text{inst}
  =
    c_d g_\xi
    \frac{(1+\mu)^{3/2+d/16}}{2^{13d/16-3/2}\mu^{3/2+3d/16}}
    \eta^{d/8}
  \epsilon^{(8-d)/4}.
\end{equation}
The instanton size $L_E$ [see Eq.\ (\ref{LE})] taken at the energy $E_*$ corresponding to $\epsilon_*$, turns out to be of the same order as $L_0$. Therefore the upper-bound condition $\epsilon< \epsilon_*$, required for the validity of the universal description, implies that the instanton size is larger than the localization length of the single-impurity bound state.

It is instructive to evaluate the instanton action (\ref{subgap-band-LO}) at $\epsilon \sim (E_{g2}-E_{g1})/2\Delta$,
which corresponds to the half-width of the impurity band. Parametrically, this energy scale coincides with $\epsilon_*$, and the action can be estimated as
\begin{equation}
  \mathcal S_\text{inst}(\epsilon_*)
  \sim \overline n_s V_\xi \left( \frac{1+\mu}{\mu} \right)^{d/4}.
\end{equation}
This action should be large in order for the present theory to be applicable
at such energies.
If the sample is thicker than $\xi_0$ (three-dimensional case, $d=3$), this condition is equivalent to the requirement
that there should be many magnetic impurities
within the localization volume, $L_0^3$,
of the single-impurity bound state [see Eq.\ (\ref{L0})].
Physically, this means good overlap of the localized impurity states, which is required for the formation of a well-defined impurity band.

\subsubsection{Smearing of the continuum gap edge ($E_{g3}$)}
\label{SSS:Eg3}

Finally, we address fluctuation smearing of the edge of the continuum spectrum, $E\to E_{g3}-0$.
In the limit (\ref{etasmall}), the third, largest root of Eq.\ (\ref{psig}) is given by
$e^{\psi_{g3}} \approx 4\mu/\eta$, and the mean-field spectrum edge is
\be
\label{Eg3}
  \frac{E_{g3}}{\Delta} \approx 1 + \frac{\eta^2}{8\mu^2}
\ee
(note that $E_{g3}$ is slightly higher than $\Delta$; this fact can be viewed as a result of level repulsion between the impurity band and the continuum). Then
$F''(\psi_{g3}) \approx -\eta/2\mu$.

In the zero-temperature limit, the effective ROP correlator
(\ref{f=f+f}) is given by
\be
\label{DOS_LOmu}
  \frac{f(0)}{\Delta^2 V_\xi}
  \approx
  \frac{\eta}{2g_\xi\mu^3}
  \left(
    - \eta
  + \frac{2\pi\mu^2}{1+\mu+\sqrt{1-\mu^2}}
  \right)^2 ,
\ee
where the two terms in the brackets correspond to the
contribution of $C(0)$ and $C_\Delta(0)$,
respectively.
At not too small $\mu$, $C_\Delta(0)$ dominates,
whereas $C(0)$ becomes the leading contribution in a very
narrow region $\mu^{4/3}\ll \eta^{2/3} \ll \mu \ll 1$,
see Fig.~{\ref{F:f/f-Eg}}(c).

The crossover energy following from Eq.\ (\ref{DOS_LOmu}) is
\begin{equation} \label{eps*contedge}
  \epsilon_*
  =
  \frac 1{2\mu^2 \eta^2}
  \left(
    \eta
  - \frac{2\pi\mu^2}{1+\mu+\sqrt{1-\mu^2}}
  \right)^4 .
\end{equation}

Equation (\ref{DOS_LOmu}) determines the main asymptotics of the DOS tail at $\epsilon\ll \epsilon_*$:
\begin{equation}
  \mathcal S_\mathrm{inst}
  =
  2^{4-d/4} c_d \frac{\Delta^2 V_\xi}{f(0)} \epsilon^{(8-d)/4} .
\end{equation}

The far asymptotics at $\epsilon\gg \epsilon_*$ is governed by the action
\begin{equation}
  \mathcal S_\mathrm{inst}
  =
  2^{(18-d)/4} s_d g_\xi \left( \frac{\mu}\eta \right)^2 \epsilon^{(6-d)/4}.
\end{equation}

The outcome of the upper-bound applicability conditions for $\epsilon$, formulated in Sec.~\ref{SS:limits}, depends on the relation between the two terms in the brackets of Eqs.\ (\ref{DOS_LOmu}) and (\ref{eps*contedge}).

In the regime  $\mu^{4/3}\ll \eta^{2/3} \ll \mu \ll 1$ [very narrow green strip in Fig.~\ref{F:f/f-Eg}(c)], where the first term in the brackets of Eqs.\ (\ref{DOS_LOmu}) and (\ref{eps*contedge}) dominates, we find that it is sufficient to require $\epsilon\ll \epsilon_2$, while $\epsilon_2 \sim \epsilon_* \approx \eta^2 /2 \mu^2$.
This means that our results based on the universal description (see Sec.~\ref{SS:universal}) are valid only in the regime of the main asymptotics, $\epsilon< \epsilon_*$. Note that $\epsilon_2$ in this limit is of the same order as the dimensionless distance between the gap edge $E_{g3}$ and $\Delta$, see Eq.\ (\ref{Eg3}).

At the same time, the dimensionless energy scale $\eta^2/\mu^2$ not only determines the difference between $E_{g3}$ and $\Delta$, but also sets the width of the DOS ``coherence peak'' above $E_{g3}$. In the regime $\mu^{4/3}\ll \eta^{2/3} \ll \mu \ll 1$, this scale coincides with $\epsilon_*$, and making a shift of this order from $E_{g3}$ into the subgap region, we find $\mathcal S_\mathrm{inst} (\epsilon\sim\epsilon_*) \sim \overline n_s A_{3-d} \xi_0^d \epsilon_*^{-d/4}$. The requirement $\mathcal S_\mathrm{inst} \gg 1$ at such energies then implies that the number of magnetic impurities within the instanton volume is large.

In the regime $\eta^{2/3} \ll \mu^{4/3}$,
where the second term in the brackets of Eqs.\ (\ref{DOS_LOmu}) and (\ref{eps*contedge}) dominates [left side of the pink region in Fig.~\ref{F:f/f-Eg}(c)], we find that the most restrictive condition is $\epsilon\ll \epsilon_4$, while
\begin{equation}
\epsilon_* / \epsilon_4 \sim \left( \mu^2 / \eta \right)^6 \gg 1.
\end{equation}
This means that the universal description breaks down at $\epsilon\sim\epsilon_4\ll\epsilon_*$,
where the subgap states are still due to fluctuations of magnetic disorder.

Finally, if $\eta \sim \mu^2$ so that the two terms in the brackets of Eqs.\ (\ref{DOS_LOmu}) and (\ref{eps*contedge}) nearly compensate each other [red line in Fig.~\ref{F:f/f-Eg}(c)], we find that the most restrictive condition is $\epsilon\ll \epsilon_3$, while $\epsilon_* \ll \epsilon_3$.
This means that the universal description applies both for the main asymptotics of the tail due to magnetic disorder (at $\epsilon< \epsilon_*$) and for its far asymptotics due to
potential disorder (at $\epsilon> \epsilon_*$).

\section{Discussion}
\label{S:discussion}

A number of experimental techniques can be used to probe peculiarities of the DOS in superconductors with magnetic impurities.
The gap suppression, predicted by the AG theory \cite{AG}, was verified by means of tunneling between normal and superconducting electrodes \cite{Exp-Woolf,Exp-Edelstein}.
The impurity band was investigated by tunneling experiments with alloys (such as PbMn) \cite{Exp-Bauriedl} and normal metal--superconductor bilayers \cite{Exp-Dumoulin}, and also by means of thermal transport in superconducting films \cite{Exp-Ginsberg}. The observation of discrete levels associated with a separate magnetic impurity was achieved in the scanning tunneling microscopy experiments \cite{Yazdani1997,Ji2008,Franke2011,Roditchev2015}.

The width $\Gamma_\mathrm{tail}$ of the DOS tail that needs to be experimentally resolved, according to our predictions, depends on the particular spectral edge.
In order to get a feeling of possible numbers, we consider the limit of weak magnetic impurities and weak spin-flip scattering, $\mu\ll \eta^{2/3}\ll 1$, with only one spectrum edge, $E_{g1}$ (see Sec.~\ref{SS:Silva}). Rewriting the main asymptotics of the DOS tail given by Eq.~(\ref{Sinst-LO}) in the form
\begin{equation}
\left< \rho(E) \right> \propto \exp\left[ - \left( \frac{E_g-E}{\Gamma_\mathrm{tail}} \right)^{(8-d)/4} \right],
\end{equation}
we find the width of the tail,
\begin{equation}
\frac{\Gamma_\mathrm{tail}}\Delta = \left( \frac{6^{d/4} \mu \eta^{1/3}}{2c_d g_\xi} \right)^{4/(8-d)}.
\end{equation}

For a small superconducting grain ($d=0$), with the parameters $\mu = 0.1$, $\eta = 0.2$, $g_\xi =10$ as an example, we obtain $\Gamma_\mathrm{tail} / \Delta \approx 0.2$.
We therefore expect that predicted tails of the DOS can be measured with the help of modern experimental techniques.
In order to distinguish the edge smearing on top of the thermal broadening, the temperature should be lower than $\Gamma_\mathrm{tail}$.

\section{Conclusions}
\label{S:concl}

We have calculated the subgap DOS tails in diffusive superconductors with pointlike magnetic impurities of arbitrary strength described by the Poissonian statistics. Hard spectral gaps obtained in the mean-field approximation are smeared due to rare fluctuations producing localized quasiparticle states in the classically forbidden region. The central question then is to identify the fluctuator responsible for the formation of the subgap states. In the present problem, there are two types of such fluctuators: (i) random potential leading to mesoscopic fluctuations (potential disorder), and (ii) concentration of magnetic impurities (nonpotential disorder).

In the framework of the replica sigma-model approach, the smearing of the hard mean-field gaps and the appearance of the tail states is described by instantons with the broken replica symmetry. In the vicinity of $E_{gi}$, the general system of instanton equations (\ref{SP-mostgeneral}) can be simplified, taking the universal form (\ref{SP12}) typical for the ROP model.
Following the general analysis of the ROP model \cite{SF}, we conclude that the competition between potential and nonpotential disorder is controlled by the parameter $\epsilon_*$ [given by Eq.\ (\ref{epsilon*})]:
close to the edge, at $\epsilon\ll\epsilon_*$, the subgap states are due to fluctuations of nonpotential disorder (LO regime), whereas the far asymptotics of the DOS, at $\epsilon\gg\epsilon_*$, is due to fluctuations of potential disorder \cite{MS}.
In both regimes, we determine the subgap DOS with the exponential accuracy
[Eqs.\ (\ref{subgap-LO}) and (\ref{subgap-MS})] by
generalizing previous results
\cite{LO_1971,LamacraftSimons,MeyerSimons2001,MS,SF}
to the case of an arbitrary function $F(\psi)$, which determines the mean-field DOS [see Eqs.\ (\ref{rho-via-psi}) and (\ref{F(psi)=0})].

In deriving the effective ROP correlation function $f(q)$, we obtain that fluctuating concentration of magnetic impurities (nonpotential disorder) affects the DOS in two ways: directly, via fluctuations of the pair-breaking parameter [see Eq.\ (\ref{C-eta-0})], and indirectly, via induced fluctuations of the order-parameter field [see Eq.\ (\ref{C-Delta-0})].
In Sec.~\ref{S:Limiting}, we demonstrate that depending on the values of $\eta$ and $\mu$ and on the particular edge considered, both mechanisms may either enhance or suppress each other.
Both mechanisms require a finite impurity strength and are absent in the Born limit ($\mu\to 0$).
In the latter case, magnetic disorder leads to the DOS smearing through the excitation of the triplet modes, rendering the effective ROP parameter $f(0)$ extremely small \cite{SF}. On the contrary, for not very weak magnetic impurities [see Eq.\ (\ref{cond*})], nonpotential disorder is not that weak and can effectively compete with potential disorder, in accordance with the general phenomenology of the ROP model.

Our present analysis also unveils the limits of the universal description based on the ROP model. With the function $F(\psi)$ [Eq.\ (\ref{F(psi)})] parametrized by the two parameters $\eta$ and $\mu$, it is possible to have a situation when the perturbative expansion in $\epsilon$ breaks down at some $\epsilon_i$ smaller than $\epsilon_*$ (this is realized, e.g., for the edge $E_{g3}$ at $\eta^{2/3}\ll\mu^{4/3}$, see Sec.~\ref{SSS:Eg3}).
In that case, at $\epsilon\sim\epsilon_i$ the LO-type behavior of the DOS tail crosses over to a different stretched-exponential behavior due to fluctuations of the same nonpotential disorder, but with different nonlinear terms.

Our reduction to the effective ROP model was performed under assumption of an arbitrary function $F(\psi)$. Therefore it may be used for the analysis of other problems when the hard gap in superconducting systems is smeared by disorder. However, each time the effective correlation function $f(0)$ should be recalculated independently.

Finally, we emphasize that our treatment is limited to the three- and zero-dimensional instanton geometries. Though our reduction to the effective ROP model is valid for any dimensionality $d$, formation of the subgap states in the ROP model in the one- and two-dimensional cases should be reconsidered due to the presence of multiple instanton solutions \cite{STS}.

\acknowledgments

We thank I.~S.\ Burmistrov, M.~V.\ Feigel'man, and P.~M.\ Ostrovsky
for stimulating discussions.
This work was supported in part by
RFBR Grant No.\ 13-02-01389.

\appendix

\section{Mean-field spectrum edges}
\label{A:lines}

Here, we discuss the mean-field spectrum edges in the superconductor with magnetic impurities. As can be seen from Figs.~\ref{F:etamudiag} and~\ref{F:egs},
there can be up to three different spectrum edges at the same time. In the case of Fig.~\ref{F:etamudiag}(d), we denote the lower and the upper edges of the impurity band by $E_{g1}$ and $E_{g2}$, respectively, while the lower edge of the continuum is denoted by $E_{g3}$.
Merging of the impurity band with the continuum, as shown in the case \ref{F:etamudiag}(b), implies merging of $E_{g2}$ and $E_{g3}$, so that there is only one spectrum edge, $E_{g1}$, left (the AG regime).
In the gapless regime \ref{F:etamudiag}(a), $E_{g1}$ turns to zero and this spectrum edge disappears as well. Alternatively, closer to the unitary limit ($\mu\to 1$), $E_{g1}$ can turn to zero while the impurity band is still present [case \ref{F:etamudiag}(c)], then there are two spectrum edges, the upper edge of the impurity band, $E_{g2}$, and the lower edge of the continuum, $E_{g3}$.

In order to find the spectrum edges $E_g$ (here $E_g$ represents any of the spectrum edges discussed above), we can start from Eq.\ (\ref{F(psi)=0}) to express $E(\psi)$. Then, changing $\psi$ from $0$ to $\infty$ along the real axis, and keeping only (physically relevant) increasing sections of the $E(\psi)$ curve, we can find the domains of energy, corresponding to zero DOS (the DOS is zero if a given energy corresponds to a real $\psi$) \cite{Shiba,Rusinov}.
Requiring $E'(\psi_g)=0$, we find an implicit equation for the angle $\psi_g$:
\be
  \frac{1+\mu(\cosh 2\psi_g -2)}{(1-\mu\cosh 2\psi_g)^2} \cosh^3 \psi_g
  = \frac{1}{\eta},
\label{psig}
\ee
which determines the spectrum edge(s) $E_g$:
\be
  \frac{E_g}\Delta = \tanh \psi_g -\eta \frac{\sinh \psi_g}{1-\mu \cosh 2\psi_g}.
\label{Eg}
\ee
Depending on the values of $\eta$ and $\mu$, the number of solutions to Eq.\ (\ref{psig})
varies from 0 to 3, corresponding to the regimes (a)--(d) in Fig.~\ref{F:etamudiag}.

The solid blue line in Fig.~\ref{F:etamudiag} corresponds to the appearance of finite DOS at $E=0$ (i.e., vanishing
of $E_{g1}$). This implies $\psi_g =0$, and Eq.\ (\ref{psig}) immediately yields a simple form $\mu = 1-\eta$ for this line.

The dashed red line corresponds to the moment of merging of the impurity band with the continuum. Two spectrum edges [corresponding to solutions of Eq.\ (\ref{psig})], $E_{g2}$ and $E_{g3}$, disappear at once, and this situation is described by equations $E'(\psi_g) = E''(\psi_g) =0$, which leads to the following analytical expression:
\begin{equation}
\label{eta-dashed-line}
  \eta
  =
  \frac{2 (2\mu)^{3/2} \bigl( 1+\sqrt{3-2\mu^2} \bigr)^2}
  {\bigl( 3+\mu + 2\sqrt{3-2\mu^2} \bigr)^{3/2} \bigl( 2-\mu +\sqrt{3-2\mu^2} \bigr)} .
\end{equation}
At small $\eta$ and $\mu$, corresponding to the lower left
corner of the diagram separating the regimes (b) and (d), this line behaves as $\mu\propto\eta^{2/3}$.
In the upper right corner, it terminates at the point $(\eta,\mu)=(4/3^{3/2},1)$.

\section{Order parameter fluctuations due to magnetic impurities}
\label{A:Delta}

\subsection{General expression for $C_\Delta(q)$}

Formation of the bound state on a single magnetic impurity is accompanied by
the suppression of the order parameter in the vicinity of the impurity \cite{Rusinov}.
The spatial scale of this suppression is given by the coherence length
(either dirty or clean, see Appendix \ref{AA:suppression}).
For many impurities, $\Delta(\br)$ becomes a random field related to
the density of magnetic impurities $n_s(\br)$ by means of Eq.\ (\ref{Delta-ns}).
The corresponding kernel $C_\Delta(q)$ is evaluated in this Appendix.

We start our consideration with the action (\ref{SSS}),
where the magnetic contribution $\mathcal S_\text{mag}$ [Eq.\ (\ref{Smag-via-ns})]
should be decomposed into the average $\mathcal S_\text{mag}^{(1)}$ [Eq.\ (\ref{Smagn1})]
and fluctuating component
$\delta\mathcal S_\text{mag} = \mathcal S_\text{mag} - \mathcal S_\text{mag}^{(1)}$
proportional to $\delta n_s$:
\begin{equation} \label{Stot}
\mathcal S = \mathcal S_\Delta + \mathcal S_D + \mathcal S_\text{mag}^{(1)}
+ \delta\mathcal S_\text{mag}[\delta n_s(\mathbf r)].
\end{equation}
The first three terms have the homogeneous saddle-point solution $\theta_{0\varepsilon}$, $\Delta_0$.
Defining fluctuation $\theta_{1\varepsilon}$ and $\Delta_1$ around the saddle-point solutions according to
\begin{subequations}
\begin{align}
\theta_\varepsilon^a(\mathbf r) &= \theta_{0\varepsilon} + \theta_{1\varepsilon}^a(\mathbf r),
\\
\Delta^a(\mathbf r) &= \Delta_0 + \Delta_1^a(\mathbf r),
\end{align}
\end{subequations}
we want to study the response of the order parameter to a particular configuration of $\delta n_s(\mathbf r)$. The inhomogeneous response $\Delta_1^a(\mathbf r)$ arises due to the last term in Eq.\ (\ref{Stot}). The ``bare'' action $\mathcal S_\Delta + \mathcal S_D + \mathcal S_\text{mag}^{(1)}$, being expanded with respect to fluctuations, produces the saddle-point value and the following second-order contribution (below, we suppress the 0 subscript of $\theta_0$ and $\Delta_0$ for brevity):
\begin{multline}
\label{S02}
  \mathcal S_0^{(2)} = \frac\nu{T} \sum_a \int d\mathbf r \frac{\left( \Delta_1^a(\mathbf r) \right)^2}\lambda
\\
+ \frac{\pi\nu}2 \sum_{\varepsilon,a}
  \int d\mathbf r\, d\mathbf r'\,
  \theta_{1\varepsilon}^a(\mathbf r) \left( \Pi_{\varepsilon} \right)^{-1} \theta_{1\varepsilon}^a(\mathbf r')
\\
-2\pi\nu \sum_{\varepsilon,a} \int d\mathbf r\, \Delta_1^a(\mathbf r) \theta_{1\varepsilon}^a(\mathbf r) \cos\theta_\varepsilon.
\end{multline}
Here, $\Pi_{\varepsilon}$ is the cooperon propagator (corresponding to variations of the
spectral angle $\theta$) on top of the superconducting state with magnetic impurities. In the momentum representation, it has the following form:
\begin{equation}
\label{Pi-def}
  \Pi_{\varepsilon}(q) = \frac{1}{Dq^2+ 2\mathfrak{E}(\varepsilon)},
\end{equation}
where
\begin{equation}
\label{mathfrakE}
  \mathfrak{E}(\varepsilon)
  =
  \varepsilon\cos\theta_\varepsilon + \Delta\sin\theta_\varepsilon
  +
  \Delta \eta \frac{\mu+\cos2\theta_\varepsilon}{(1+\mu\cos2\theta_\varepsilon)^2} .
\end{equation}
The homogeneous saddle-point equation for the spectral angle $\theta_\varepsilon$ reads
\begin{equation} \label{Usadeltheta}
-\frac\varepsilon\Delta \sin\theta_\varepsilon + \cos\theta_\varepsilon - \frac\eta 2 \frac{\sin 2\theta_\varepsilon}{1+\mu\cos 2\theta_\varepsilon} =0
\end{equation}
[this is the Matsubara-frequency version of Eq.\ (\ref{F(psi)=0}), written in terms of $\theta = \pi/2 +i\psi$].

In order to find the response of $\Delta_1(\mathbf r)$ to the field $\delta n_s(\br)$,
we expand the term $\delta\mathcal S_\text{mag}$
to the first order in $\theta_1$ and obtain
\begin{multline} \label{Delt1ns}
  \Delta_1^a(\mathbf r)
  =
  - \sum_{\varepsilon,b} \frac{\mu \sin 2\theta_{\varepsilon}}{1+\mu \cos 2\theta_{\varepsilon}}
\\
\times \int d\mathbf r' \left< \Delta_1^a(\mathbf r) \theta_{1\varepsilon}^b (\mathbf r') \right>_0 \delta n_s(\mathbf r').
\end{multline}
The average over the Gaussian action (\ref{S02}) has the form
\begin{equation}
  \left< \Delta_1^a \theta_{1\varepsilon}^b \right>_{0\mathbf q}
  =
  \delta_{ab} \frac T\nu L(q) \Pi_\varepsilon(q) \cos\theta_{\varepsilon} ,
\end{equation}
where $L(q)$ is the static longitudinal propagator of superconducting fluctuations:
\be
\label{L(q)}
  L^{-1}(q)
  =
  \pi T \sum_\varepsilon
  \left[ \frac{\sin\theta_\varepsilon}{\Delta} - \frac{2\cos^2 \theta_\varepsilon}{Dq^2+ 2\mathfrak{E}(\varepsilon)}
  \right]
  .
\ee
Equation (\ref{Delt1ns}) then yields a replica-symmetric response
(\ref{Delta-ns}) with the kernel
\begin{equation}
\label{Delta1nsfinal}
  C_\Delta(q)
  =
  \frac{\mu}\nu L(q) T \sum_\varepsilon \frac{\sin 2\theta_\varepsilon \cos\theta_\varepsilon}{1+\mu \cos 2\theta_\varepsilon}
  \Pi_\varepsilon(q) .
\end{equation}

\subsection{$\Delta(\br)$ suppression near a single magnetic impurity}
\label{AA:suppression}

As a byproduct of our consideration, we can find
the suppression of $\Delta$ in the vicinity of a single magnetic impurity. For that, in Eq.\ (\ref{Delta1nsfinal}) we should take $L(q)$ and $\Pi_\varepsilon(q)$ on the background of purely potential scattering ($\overline n_s=0$, $\eta=0$) with
$\cos\theta_\varepsilon = \varepsilon/\mathfrak{E}(\varepsilon)$,
$\sin\theta_\varepsilon = \Delta/\mathfrak{E}(\varepsilon)$,
and $\mathfrak{E}(\varepsilon)=\sqrt{\varepsilon^2+\Delta^2}$
[as found from Eq.\ (\ref{Usadeltheta})].
A single magnetic impurity (at $\mathbf r=0$) is described by $\delta n_s(\mathbf q)=1$,
and we get simply
\begin{equation} \label{Delta1nsfinal1imp}
\Delta_1(q)
=
-C_\Delta(q) .
\end{equation}
In the real space, the order parameter is suppressed on a length scale of the order of $\xi_0$ near the magnetic impurity.

The result (\ref{Delta1nsfinal1imp}) derived in the diffusive limit can be easily extended to the case of an arbitrary mean free path $l$ by using a more general expression
in Eq.\ (\ref{Delta1nsfinal}) for the cooperon propagator \cite{LO_1971}:
\begin{equation}
\label{Pi-l}
\Pi_{\varepsilon}^{(l)}(q) = \frac{\tau\arctan\left( \frac{ql}{2\tau\sqrt{\varepsilon^2+\Delta^2} +1} \right) / ql}
{1- \arctan\left( \frac{ql}{2\tau\sqrt{\varepsilon^2+\Delta^2} +1} \right) / ql}.
\end{equation}
In particular, in the ballistic limit at Matsubara energies and real $q$,
\begin{equation}
  \Pi_{\varepsilon}^{(l=\infty)}(q)
  =
  \frac{1}{v_F q}
  \arctan\left( \frac{v_F q}{2\sqrt{\varepsilon^2+\Delta^2}} \right),
\end{equation}
where $v_F$ is the Fermi momentum,
and we readily reproduce the result by Rusinov \cite{Rusinov}.
The spatial scale of the order parameter suppression is then the clean coherence length $v_F/\Delta$.

\subsection{Zero-temperature limit for $C_\Delta(0)$}

The general expression for the kernel $C_\Delta(q)$ is given by Eq.\ (\ref{Delta1nsfinal}).
The value of $C_\Delta(0)$ at zero momentum can be easily evaluated
at $T=0$ by switching from integration over $\varepsilon$
to integration over $\theta_\varepsilon$ with the help of the relation
\begin{equation}
  d\varepsilon/d\theta_\varepsilon = - \mathfrak{E}(\varepsilon)/\sin\theta_\varepsilon,
\end{equation}
derived from Eqs.\ (\ref{mathfrakE}) and (\ref{Usadeltheta}) (note also that the spectral angle $\theta_\varepsilon$ is real in the Matsubara technique).
Then, $T\sum_\varepsilon (\dots)$ in Eq.\ (\ref{Delta1nsfinal}) can be calculated as
\begin{equation}
\label{Y}
  \frac{1}{\pi} \int_0^{\theta_*}
  \frac{d\theta \cos^2\theta}{1+\mu \cos 2\theta}
  =
  \frac{\Upsilon(\mu,\theta_*)}{\pi} ,
\end{equation}
where we introduced the function
\begin{equation}
\label{Upsilon}
  \Upsilon(\mu,\theta_*)
  =
  \frac{\theta_*-\sqrt{\frac{1-\mu}{1+\mu}}\arctan\left(\sqrt{\frac{1-\mu}{1+\mu}}\tan\theta_* \right)}{2\mu} ,
\end{equation}
and $\theta_*$ is the value of the spectral angle at $\varepsilon=0$:
\be
\label{theta-E0}
  \theta_*
  =
  \begin{cases}
    \pi/2, & \eta+\mu<1 , \\
    \arcsin\frac{\sqrt{\eta^2+8\mu^2+8\mu}-\eta}{4\mu}, & \eta+\mu>1 .
  \end{cases}
\ee
Analogously, from Eq.\ (\ref{L(q)}) we obtain an expression for the fluctuation propagator
in the limit of $T=0$, $q=0$:
\be
\label{L00}
  L^{-1}(0)
  =
  1 - \eta \Upsilon(\mu,\theta_*)
.
\ee

Finally, substituting everything into Eq.\ (\ref{Delta1nsfinal})
and using Eq.\ (\ref{eta-def}), we obtain
\begin{equation}
\label{C_Delta(0)-T0}
  C_\Delta(0)
  =
  \frac{\Delta}{\overline{n}_s}
  \frac{\eta\Upsilon(\mu,\theta_*)}{1 - \eta \Upsilon(\mu,\theta_*)}
  .
\end{equation}
In the gapped phase [$\eta+\mu<1$; regions (b) and (d) in Fig.~\ref{F:etamudiag}],
$\theta_*=\pi/2$ and
$\Upsilon(\mu,\theta_*) = (\pi/2)/(1+\mu+\sqrt{1-\mu^2})$.
With the help of Eq.\ (\ref{gxi}), Eq.\ (\ref{C_Delta(0)-T0}) can be reduced to Eq.\ (\ref{C-Delta-0}).


\begin{thebibliography}{99}
\bibitem{mu-comment}
The parameter $\mu$ can be expressed via the scattering phase
on the magnetic impurity as $\mu=\sin^2\delta/(1+\cos^2\delta)$,
where $\delta=\delta_0^+-\delta_0^-$, and $\delta_0^\pm$
is the isotropic ($l=0$) electron scattering phase with the spin parallel (antiparallel) to the spin of the impurity \cite{Rusinov}.
\bibitem{com-L0}
For an arbitrary strength of potential scattering,
decay of the state localized on a single magnetic impurity
is determined by the pole of the cooperon propagator (\ref{Pi-l}):
$\Pi_{iE_0}^{(l)}(i/L_0)=\infty$. In the ballistic case,
one obtains $L_0 = (v_F/2\Delta) (1-E_0^2/\Delta^2)^{-1/2}$,
in agreement with Ref.~\cite{Rusinov}.
\bibitem{com-4}
In Ref.~\cite{SF}, the coefficient in front of the sigma-model action (\ref{S0}) was overestimated by the factor of 2. At the same time, the anzats similar to (\ref{psi-RSB}) did not respect the symmetry constraint (\ref{Q-constraint}) and neglected the contribution from the energy $-\varepsilon_0$. These two errors compensated each other, leading to the correct instanton action.
\bibitem{l-com}
The fact that we reproduce the results in the Born limit ($\overline n_s\to\infty$)
indicates that the actual condition might be much less restictive.
\bibitem{xi-com}
Strictly speaking, in the presence of magnetic disorder, the effective
coherence length, $\xi$, differs from the nonmagnetic value given
by Eq.\ (\ref{xi-def}). It can be estimated from the expression (\ref{Pi-def})
for the cooperon as $\xi(T)=\sqrt{D/2\mathfrak{E}(\pi T)}$,
where $\pi T$ is the lowest Matsubara energy.
At zero temperature, $\mathfrak{E}(0)=\Delta(1-\mu-\eta)/(1-\mu)$
in the gapped phase [regions (b) and (d) in Fig.~\ref{F:etamudiag}], while in the gapless
phase [regions (a) and (c)] $\mathfrak{E}(0)$ can be obtained
from Eqs.\ (\ref{mathfrakE}) and (\ref{theta-E0}).
\bibitem{com-d-2}
The radial part of the Laplace operator in a $(d-2)$-dimensional space is $(\nabla^2)_\text{rad} = \partial_r^2 + (d-3) r^{-1} \partial_r$.
\bibitem{MScomment}
Our result (\ref{subgap-MS}) for the action is four times smaller than Eq.\ (15) of Ref.~\cite{MS}. At the same time, the action (13) from Ref.~\cite{MS}, written for the $d=1$ case, is also four times smaller than what follows from their Eq.\ (15).
To resolve the apparent inconsistency within Ref.~\cite{MS}, we suggest that their coefficient $a_d$ should be defined with an additional factor of 1/4. Then Eqs.\ (13) and (15) from Ref.~\cite{MS} agree with each other and with our Eq.\ (\ref{subgap-MS}).
\end{thebibliography}
\end{document}